\documentclass[preprintnumbers,article,amsmath,amssymb,floatfix,10pt,prd,twocolumn,
superscriptaddress,nofootinbib]{revtex4-2}
\usepackage{bm}
\usepackage{amsfonts}
\usepackage{latexsym}
\usepackage[utf8]{inputenc}
\usepackage{graphicx}
\usepackage{amsmath}
\usepackage{dirtytalk}
\usepackage{palatino}
\usepackage{mathpazo}
\usepackage{textcomp}
\usepackage{float}
\usepackage{booktabs}
\usepackage{dcolumn}
\usepackage{ragged2e}
\usepackage{hyperref}
\hypersetup{colorlinks,citecolor=blue}
\hypersetup{colorlinks=true,linkcolor=red,filecolor=magenta,    urlcolor=blue}
\usepackage{amsmath}
\usepackage{xcolor}
\usepackage{orcidlink}
\usepackage{epsfig}
\usepackage{caption}
\usepackage{subcaption}
\usepackage{commath}
\def\jnl@style{\it}
\def\aaref@jnl#1{{\jnl@style#1}}

\def\aaref@jnl#1{{\jnl@style#1}}

\def\aj{\aaref@jnl{AJ}}                   % Astronomical Journal
\def\apj{\aaref@jnl{ApJ}}                 % Astrophysical Journal
\def\apjl{\aaref@jnl{ApJ}}                % Astrophysical Journal, Letters
\def\apjs{\aaref@jnl{ApJS}}               % Astrophysical Journal, Supplement
\def\apss{\aaref@jnl{Ap\&SS}}             % Astrophysics and Space Science
\def\aap{\aaref@jnl{A\&A}}                % Astronomy and Astrophysics
\def\aapr{\aaref@jnl{A\&A~Rev.}}          % Astronomy and Astrophysics Reviews
\def\aaps{\aaref@jnl{A\&AS}}              % Astronomy and Astrophysics, Supplement
\def\mnras{\aaref@jnl{Mon.~Not.~Roy.~Astron.~Soc.}}             % Monthly Notices of the RAS
\def\prd{\aaref@jnl{Phys.~Rev.~D}}        % Physical Review D
\def\prc{\aaref@jnl{Phys.~Rev.~C}}  % Physical Review C
\def\prl{\aaref@jnl{Phys.~Rev.~Lett.}}    % Physical Review Letters
\def\qjras{\aaref@jnl{QJRAS}}             % Quarterly Journal of the RAS
\def\skytel{\aaref@jnl{S\&T}}             % Sky and Telescope
\def\ssr{\aaref@jnl{Space~Sci.~Rev.}}     % Space Science Reviews
\def\zap{\aaref@jnl{ZAp}}                 % Zeitschrift fuer Astrophysik
\def\nat{\aaref@jnl{Nature}}              % Nature
\def\aplett{\aaref@jnl{Astrophys.~Lett.}} % Astrophysics Letters
\def\apspr{\aaref@jnl{Astrophys.~Space~Phys.~Res.}} % Astrophysics Space Physics Research
\def\physrep{\aaref@jnl{Phys.~Rep.}}      % Physics Reports
\def\physscr{\aaref@jnl{Phys.~Scr}}       % Physica Scripta
\def\commat{\aaref@jnl{Comm.~Math.~Phys.}}              % Communications in Mathematical Physics
\def\science{\aaref@jnl{Science}}               % Science
\def\cqg{\aaref@jnl{Classical Quant.~Grav.}}            % Classical and Quantum Gravity
\def\jpcs{\aaref@jnl{JPCS}}                                     % Journal of Physics Conference Series
\def\ijmpd{\aaref@jnl{Int.~J.~Mod.~Phys.~D}}                    % International Journal of Modern Physics D
\def\grg{\aaref@jnl{Gen.~Relat.~Gravit.}}               % General Relativity and Gravitation
\def\rpp{\aaref@jnl{Rep.~Prog.~Phys.}}          % Reports on Progress in Physics
\def\npa{\aaref@jnl{Nucl.~Phys.~A}}        % Nuclear Physics A
\def\lrr{\aaref@jnl{Living Rev.~Rel.}}                   % Living reviews in relativity
\def\jcap{\aaref@jnl{J.~Cosmology Astropart.~Phys.}}    % Journal of cosmology and astroparticle physics
\def\rmp{\aaref@jnl{Rev.~Mod.~Phys.}}   %Reviews of modern physics
\def\epjc{\aaref@jnl{Eur.~Phys.~J.~C}} 
\def\plb{\aaref@jnl{~Phy.~Lett.~B}} 
\def\mpla{\aaref@jnl{Mod.~Phy.~Lett.~A}} 
\def\arxiv{\aaref@jnl{arxiv.org}}

\allowdisplaybreaks[1]
\begin{document}
%\color{red}
\color{black}       %% For one column

\title{Resolving FLRW cosmology through effective equation of state in $f(T)$ gravity}
%\end{document}

\author{S. R. Bhoyar\orcidlink{0000-0001-8427-4540}}
\email{drsrb2014@gmail.com}
\affiliation{Department of Mathematics, Phulsing Naik Mahavidyalaya Pusad-445216 Dist. Yavatmal (India).}

\author{Yash B. Ingole\orcidlink{0009-0006-7208-1999}}
\email{ingoleyash01@gmail.com}
\affiliation{Department of Mathematics, Phulsing Naik Mahavidyalaya Pusad-445216 Dist. Yavatmal (India).}
%
%%%%%%%%%%%%%%%%%%%%%%%%%%%%%%%%%%%%%  DATE  %%%%%%%%%%%%%%%%%%%%%%%%%%%%%%%%%%%%

\begin{abstract}

\textbf{Abstract:} This article explores the cosmological scenario of the universe in the context of the $f(T)$  power law model, where $T$ represents the torsion scalar. To obtain the deterministic solution of the field equations, we parameterized the effective equation-of-state with two parameters $m$ and $k$ as suggested by A. Mukherjee in a flat FLRW environment. We impose constraints on the free parameters of the derived solution by utilizing MCMC analysis, assuming the $CC, Pantheon+SH0ES$, and $CC+Pantheon+SH0ES$ as data samples. We explore the dynamics of cosmological parameters. The evolutionary profile of the deceleration parameter exhibits the transition %from the decelerated 
to the accelerated phase. The effective equation-of-state parameter indicates that the model remains in the quintessence era and gradually becomes the Einstein-de Sitter model. In addition, we explore the jerk, snap, and lerk parameters. Furthermore, the $Om(z)$ diagnostic shows that the model has a consistent positive slope across the entire evolution, but soon resembles the standard $\Lambda$CDM model. Finally, we conclude that the power law model of the  $f(T)$ gravity in the framework of the FLRW universe aligns more closely with the $\Lambda$CDM model for given observational data.\\

\end{abstract}

\maketitle

\date{\today}

%%%%%%%%%%%%%%%%%%%%%%%%%%%%%%%%%%%%%%%%%%%%%%%%%%%%%%%%%%%%%%%%%%%%%%%%
%%%%%%%%%%%%%%%        Introduction        %%%%%%%%%%%%%%%%%%%%%%%%%%%%%
%%%%%%%%%%%%%%%%%%%%%%%%%%%%%%%%%%%%%%%%%%%%%%%%%%%%%%%%%%%%%%%%%%%%%%%%

\section{Introduction}
\label{section 1}

A wide range of cosmological observations, such as those of Type Ia Supernovae (SNeIa) \cite{riess1998observational,perlmutter1999measurements}, cosmic microwave background radiation (CMBR) \cite{spergel2003first,spergel2007three}, and large-scale structures \cite{tegmark2004cosmological,eisenstein2005detection}, have demonstrated that the universe is expanding at an accelerated rate. This surprising observation of a phenomenon raises one of the most challenging issues in modern cosmology. An unidentified so-called dark energy (DE) is the primary gradient that causes the acceleration of the universe. This exotic energy component, characterized by negative pressure, influences the universe, which is filled with cold dark matter (CDM), and drives the recent acceleration of the expansion of the universe. Over the last ten years, many DE models, such as quintessence, phantom, k-essence, tachyon, and quintom \cite{caldwell2002phantom} as well as Chaplygin gas \cite{kamenshchik2001alternative}, generalized Chaplygin gas (GCG) \cite{bento2002generalized}, holographic DE \cite{li2004model}, new agegraphic DE \cite{wei2008cosmological}, and Ricci DE \cite{gao2009holographic} have been suggested.
The characteristics of DE are currently the most challenging issues in particle physics and modern cosmology. Approximately 75\% of the total energy of the universe is thought to be covered by DE. The constant equation-of-state (EoS) parameter $\omega \approx -1$ and the vacuum energy (cosmological constant, $\Lambda$) are the most straightforward and appealing candidates for DE. It becomes difficult to reconcile the
small observational value of DE density that comes from quantum field theories; this is called the cosmological constant problem \cite{copeland2006dynamics}. However, two methods have been used to explain the universe's recent acceleration. The first method involves experimenting with the energy-momentum tensor in Einstein's field equation, which includes a scalar field. The other method involves a geometrical modification of Einstein-Hilbert action \cite{paliathanasis2016cosmological} because it describes the nature of DE as a geometrical aspect of the universe called the modified theory of gravity.

The scientific literature has proposed and discussed numerous theories on modified gravity. These theories aim to extend or alter the framework of general relativity (GR) to address unresolved issues in cosmology, such as the nature of DE and dark matter (DM), or to explain the observed acceleration of the universe's expansion without invoking these mysterious components. A top-notch theory in this regard is $f(R)$ gravity \cite{sharif2013dynamical}, which is a simple modification of general relativity that introduces an arbitrary function of the Ricci scalar $R$ into the Einstein-Hilbert action. This modified gravity theory is recognized for its ability to effectively account for cosmic acceleration and can accurately reproduce the entire cosmological history, including the behavior of the cosmological constant \cite{kadam2022teleparallel}. In an alternative approach, instead of using the curvature defined by the Levi-Civita connection, one can explore the Weitzen\"ock connection, which has no curvature but instead features torsion, this torsion is constructed entirely from products of the first derivatives of the tetrad, with no second derivatives appearing in the torsion tensor. This approach is known as teleparallelism and is closely equivalent to standard GR, differing only in boundary terms and involving total derivatives in the action. Building on the formulation of $ f(R)$ gravity, we extended the teleparallel theory by introducing an arbitrary function of the torsion scalar \( T \), resulting in what is known as \( f(T) \) gravity. In this generalized theory, the dynamics of gravity are governed by the curvature of spacetime, as in traditional GR, but by the torsion associated with the Weitzenb\"ock connection. By allowing the torsion scalar \( T \) to vary according to an arbitrary function, \( f(T) \) gravity offers a new framework for exploring gravitational phenomena, potentially providing new insights into the nature of cosmic acceleration and the evolution of the universe. Additionally, compared with $f(R)$ gravity, cosmological scenarios in $f(T)$ gravity are easier to explore. Several works have been done in $f(T)$ gravity. Recently, investigations of cosmological possibilities, such as late-time cosmic acceleration \cite{bengochea2009dark,arora2021constraining,gadbail2023dark,myrzakulov2023constrained} and the inflationary model \cite{bamba2016bounce}, have been conducted. Furthermore, research has been conducted in $f(T)$ gravity on observational constraints, dynamical systems \cite{nunes2018structure,duchaniya2024attractor}, and structure creation, assuming that $f(T)$ is a power law function.

The analysis focused on scalar perturbations in \( f(T) \) gravity and their impact on the anisotropy of the cosmic microwave background (CMB).  M. Blagojevic et al. \cite{blagojevic2024lorentz} studied the Lorentz invariant defined by the coframe-connection-multiplier form of the Lagrangian in teleparallel gravity. They also provide a detailed Hamiltonian analysis showing that local Lorentz invariance is completely broken in 
$f(T)$ theories, while also addressing the number of physical degrees of freedom and confirming diffeomorphism invariance \cite{blagojevic2020local}. Ferraro et al. \cite{ferraro2015remnant} discussed how extended teleparallel gravitational theories inherit some local Lorentz invariance associated with the tetrad field. They analyze various spacetime examples and conclude that the symmetry structure is more intricate than commonly assumed. 
This research explores how modifications in teleparallel theories affect local Lorentz invariance and discusses implications for the physical interpretation of these theories.
\cite{krvsvsak2016covariant} presents a covariant formulation of $f(T)$ gravity, addressing the problems of frame dependence and the violation of local Lorentz invariance. Their approach introduces both the tetrad and the spin connection, making $f(T)$ gravity covariant and consistent, avoiding the issues seen in the pure tetrad version. Lorentz invariance in the classical formulation of $f(T)$ gravity and the covariant spin connection restores some form of Lorentz invariance, as discussed by \cite{golovnev2021issues}. However, Golovnev argues that this covariantization does not alter the core equivalence with the original formulation. \cite{bahamonde2023teleparallel} provided a broad review of teleparallel gravity, including its covariant formulations and modified versions such as $f(T)$ gravity. They highlight how covariant formulations handle local Lorentz invariance and discuss the implications for cosmology. \cite{hu2023effective} explored the strong coupling issue in $f(T)$ gravity using an effective field theory (EFT) approach. They investigated scalar perturbations and reported that while a strong coupling problem may arise, it can be avoided under certain conditions. \cite{harko2014nonminimal} introduces an extension of $f(T)$ gravity with non-minimal torsion-matter coupling. They find that this model leads to a variety of cosmological phenomena, including quintessence and phantom-like dark energy behaviors. \cite{tajahmad2017studying} investigates the addition of an unprecedented term coupling the torsion scalar with both a scalar field and its derivatives, showing that this new term behaves consistently without producing anomalies
. While comparing the modified gravitational luminosity distance and electromagnetic luminosity distance, \cite{chen2024prospects} various cosmological models in the framework of $f(T)$ gravity have been explored.

In our current study, we aim to solve FLRW cosmology via the effective EOS parameter $\omega_{eff} = -\frac{1}{1 + m(1+z)^k}$, as discussed previously \cite{mukherjee2016acceleration}, within the framework of $f(T)$ gravity theory. One of the simplest and earliest parameterizations introduced by Chevallier and Polarski, later refined by Linder, is CPL parameterization \cite{chevallier2001accelerating}. This approach expresses the effective EoS parameter, \( \omega_{eff} \), as a first-order Taylor expansion in terms of the scale factor \( a \), given by \( \omega_{eff}(a) = \omega_0 + \omega_a(1 - a) \), where $\omega_0$, and $\omega_a$ are free parameters. In terms of redshift, this becomes \( \omega_{eff}(z) = \omega_0 + \omega_a \frac{z}{1+z} \). While the CPL parameterization behaves smoothly at early times (as \( z \to \infty \)) and at present \( z = 0 \), it diverges as \( z \to -1 \) in the future. On the other hand, the parameterization we conceive in this article does not diverge for any choice of model parameters $m$ and $k$ in the future. A good parameterization should ideally remain well-behaved across all relevant redshift ranges. To gain future insights, we avoid CPL parameterization and focus on the parametric reconstruction of the effective EoS suggested by \cite{mukherjee2016acceleration}. We present a detailed cosmological model based on the power-law formulation of $f(T)$ gravity in the FLRW universe. From this, we derive the effective EOS parameter via the Friedmann equation specific to $f(T)$ gravity. To obtain an exact solution for the Hubble function, we employ a novel approach. To fine-tune the model and obtain the best-fit values for the parameters $m$, $n$, and $k$, we utilize the $\chi^2$ minimization technique. This statistical approach helps us compare the theoretical predictions with observational data and identify the parameter set that best aligns with the empirical evidence. The datasets used for this analysis include  Cosmic Chronometers ($CC$) data, the $Pantheon+SH0ES$ Supernovae dataset, and a combined analysis of both. These datasets provide a comprehensive and diverse set of observational constraints, allowing us to test and validate our cosmological model rigorously.

This article unfolds through the following sections: Section \eqref{section 2} includes an in-depth review of the modified $f(T)$ gravity, where we use its power law model in the framework of the FLRW universe. In section \eqref{section 3}, we employ 31 Hubble data points and 1701 $Pantheon+SH0ES$ data points to constrain the model parameters, and the model is compared with the standard $\Lambda$CDM model using error bar plots. In section \eqref{section 4}, we illustrate the behavior of the model with the given best-fit values of the model parameters via. the deceleration parameter, effective EoS parameter, jerk, snap, and lerk parameter. To distinguish the different DE models by analyzing the Hubble parameter, the $Om$(z) diagnostic is discussed in section \eqref{section 5}. Finally, Section \eqref{section6} provides a summary of our work and discusses the results obtained.

\section{The Mathematical Formulation of the $f(T)$ Gravity and FLRW Universe}
\label{section 2}
Teleparallelism employs a vierbein field $e_i(x^\mu)$, where $i = 0,1,2,3 $ as its dynamical entity, serving as an orthonormal basis within the tangent space at each point $x^\mu$ on the manifold $e_i.e_j=\eta_{ij}$ where, $\eta_{ij}=diag(+1,-1,-1,-1)$. The components of each vector $e_i$  can be expressed as $e_i^{\mu}$, where $\mu=0,1,2,3$ on a coordinate basis, denoted by $e_i=e_i^{\mu}\partial_\mu$. Latin indices are related to the tangent space, whereas Greek indices indicate coordinates on the manifold. The metric tensor $g_{\mu\nu}(x)$ is derived from the dual vierbein by expressing it as  $g_{\mu\nu}(x)=\eta_{ij}e{^i_\mu}(x)e{^j_\nu}(x)$. Compared with general relativity, which uses the torsionless Levi-Civita connection ${\Gamma}{^\lambda_{\mu\nu}}=\frac{1}{2} g^{\lambda\gamma}(g_{\gamma\mu,\nu}+g_{\gamma\nu,\mu}-g_{\mu\nu,\gamma})$, teleparallelism employs the curvatureless Weitzenb\"ock connection $\Hat{\Gamma}{^\gamma{}_{\mu\nu}}=e{^\alpha_i}\partial_\nu e{^i_\mu}=-e{^\alpha_i}\partial_\mu e{^i_\nu}$, which is characterized by its non-zero torsion.
\begin{equation}\label{1}
T{^\gamma{}_{\mu\nu}}=\Hat{\Gamma}{^\gamma{}_{\nu\mu}}-\Hat\Gamma{^\gamma{}_{\mu\nu}}=e{^\gamma_i}(\partial_\mu e{^i_\nu}-\partial_\nu e{^i_\mu}),
\end{equation}
Every piece of knowledge regarding the gravitational field is contained in this tensor. The TEGR Lagrangian is constructed using a torsion tensor [eqn.\eqref{1}], and its equation of motion for the vierbein leads to the Einstein equations for the metric. The teleparallel Lagrangian is given as \cite{aldrovandi2012teleparallel}

\begin{equation}\label{2}
L_T \equiv T=T{^\gamma{}_{\mu\nu}}S{_\gamma{}^{\mu\nu}},
\end{equation}
where,
\begin{equation}\label{3}
    S{_\gamma{}^{\mu\nu}}=\frac{1}{2}(K{^{\mu\nu}{}_\gamma}+\delta{^\mu_\gamma}T{^{\beta\nu}{}_\beta}-\delta{^\nu_\gamma}T{^{\beta\mu}{}_\beta}),
\end{equation}
and $K{^{\mu\nu}{}_\gamma}$ is a contortion tensor
\begin{equation}\label{4}
    K{^{\mu\nu}{}_\gamma}=-\frac{1}{2}(T{^{\mu\nu}{}_\gamma}-T{^{\nu\mu}{}_\gamma}-T{_\gamma{}^{\mu\nu}}),
\end{equation}
This equals the difference between Weitzenb\"ock and Levi-Civita connections.

A homogeneous and isotropic flat Friedmann-Lema\^itre-Robertson-Walker (FLRW) background metric is given as
\begin{equation}\label{5}
   ds^2=dt^2-a^2(t)(dx^2+dy^2+dz^2), 
\end{equation}
where $a(t)$ is the scale factor.
Let us choose the following set of diagonal tetrads
related to the metric \eqref{5}
\begin{equation}\label{6}
    e^i_\mu = diag[1,a(t),a(t),a(t)].
\end{equation}

The diagonal tetrad is often used in cosmological models such as the FLRW framework because it simplifies the field equations. In teleparallel gravity, different tetrad choices lead to different field equations and cosmological solutions. The diagonal tetrad is favored for the FLRW universe because it maintains homogeneity and isotropy, making calculations easier by reducing the equations of motion to second-order differential equations. This approach is consistent with existing research in \(f(T)\) cosmology \cite{bengochea2009dark,nunes2018structure,duchaniya2024attractor,rodrigues2012anisotropic}.

we use the simplified torsion scalar instead of calculating it directly from the eqn. \eqref{2}.
\begin{equation}\label{7}
    T=-6H^2,
\end{equation}
where, $H$ is a Hubble parameter, and $H=\frac{\Dot{a}}{a}$, where overdot $ (^.)$ represents the derivative with respect to the cosmic time $t$. This choice is based on the established result in FLRW cosmology, which links \( T \) directly to the Hubble parameter \( H \). This helps streamline the analysis, which is particularly useful when dealing with large datasets or complex parameter constraints (like in MCMC analyses) \cite{duchaniya2024attractor}.

We can extend $T$ as a function, $T+f(T)$ and the sectors of matter and the radiation must be considered. Let us implement $f(T)$  gravity within a cosmological context. The action in $ f(T)$ gravity is defined in \cite{bengochea2009dark,cai201679j6901c} as
\begin{equation}\label{8}
    S=\frac{1}{16\pi G}\int\left[T+f(T)\right]ed^4x +S_m +S_r,
\end{equation}
assuming that the matter and radiation Lagrangian correspond to perfect fluids, $\ f(T)$ denotes an algebraic function of the torsion scalar $T$, and $\sqrt{-g}=det[e{^i_\mu}]=e$ and  $G$ is the gravitational constant. The pressure and energy densities corresponding to the Lagrangian are denoted by $P_r$, $P_m$, and $\rho_r$, $\rho_m$  respectively.

The equation of motion is obtained by the functional variation of the action \eqref{8} concerning the tetrad as
\begin{widetext}
\begin{equation}\label{9}
    e^{-1} \partial_\mu(ee^\gamma_i S_\gamma{}^{\mu\nu})[1+f_T]+e^\gamma_i S_\gamma{}^{\mu\nu}\partial_\mu(T)f_{TT}-[1+f_{TT}] e^\lambda_i T^\gamma{}_\mu \lambda S_\gamma{}^\nu\mu +\frac{1}{4}e^\nu_i [T+f(T)]=4\pi G e^\gamma_i [T^{(m)}{}_\gamma{}^\nu + T^{(r)}{}_\gamma{}^\nu],
\end{equation}
\end{widetext}
where $T^{(m)}{}_\gamma{}^\nu ,  T^{(r)}{}_\gamma{}^\nu$ are the matter and radiation energy-momentum tensors, respectively. $f_T$ and $f_{TT}$ represent the first and second order derivatives of $f(T)$ with respect to the torsion scalar $ T$.

By substituting the vierbein from eqn. \eqref{6} into eqn. \eqref{9} under the conditions $( i = 0 = \mu )$ and $( i = 1 = \mu )$, we obtain
\begin{equation}\label{10}
    H^2=\frac{8\pi G}{3} (\rho_m +\rho_r)-\frac{f}{6} + \frac{T f_T}{3},
\end{equation}
\begin{equation}\label{11}
    \Dot{H}=- \frac{4\pi G (\rho_m +P_m +\rho_r +P_r)}{1+f_T+2Tf_{TT}}.
\end{equation}
Inspired by G.R. Bengochea and R. Ferraro, \cite{bengochea2009dark} we characterized the power law model of $f(T)$ gravity as
\begin{equation}\label{12}
    f(T)=\alpha (-T)^n,
\end{equation}
where $\alpha,n$ are model parameters. By examining the form of the first Friedmann equation \eqref{10}, we infer that in \( f(T) \) cosmology, an effective dark energy component emerges from gravitational origins. Specifically, we can define the effective dark energy density and pressure as follows \cite{cai201679j6901c}:
 \begin{equation}\label{13}
     \rho_{eff} \equiv \frac{1}{16\pi G}\bigl(2Tf_T-f\bigr),
 \end{equation}
\begin{equation}\label{14}
    P_{eff}\equiv\frac{1}{16\pi G}\biggl(\frac{f-f_T T +2T^2f_{TT}}{1+f_T +2Tf_{TT}}\biggr).
\end{equation}
Furthermore, the equation of continuity for DE and matter is written as,
\begin{equation}\label{15}
    \Dot{\rho}_{eff}+3H(\rho_{eff}+P_{eff})=0,
\end{equation}
\begin{equation}\label{16}
    \Dot{\rho}_m+3H(\rho_{m}+P_{m})=0.
\end{equation}

The effective EOS parameter is defined as the ratio of the pressure \eqref{14} to the energy density \eqref{13}. It is possible to define the contribution of $f(T)$ gravity in the following way:
\begin{equation}\label{17}
    \omega_{eff}=- \frac{f/T -f_T +2Tf_{TT}}{[1+f_T+2Tf_{TT}][f/T -2f_T]},
\end{equation}
From eqns. \eqref{7} and \eqref{12} the effective EoS \eqref{17} is obtained as
\begin{equation}\label{18}
    \omega_{eff} = -\frac{1-n}{1+\alpha n (2n-1) ( -6 H^2)^{n-1}}.
\end{equation}
We require one more alternative equation to solve equation \eqref{18} for $H$. Now, we assume that the effective EoS parameter is a function of the redshift $z$ to have the well-motivated parametric form given by A. Mukherjee \cite{mukherjee2016acceleration}.
\begin{equation}\label{19}
    \omega_{eff}= - \frac{1}{1+m(1+z)^k},
\end{equation}
where $m$ and $k$ are  model parameters. The exotic component known as DE is responsible for the majority of the contributions in the recent past. The effective EoS at high redshift was essentially void because the DM was pressureless. It has a negative value during the recent acceleration period that is less than $ -1/3$. These two stages of evolution can be easily accommodated by the functional form of the effective EoS \eqref{19} utilized for the current reconstruction. The values of $\omega_{eff}$ \eqref{19} tend to zero at a high redshift $z$  for positive values of the model parameters $m$ and $k$, and for $z=0$, the value of the effective EoS depends on the value of the model parameter. A positive value of the model parameters establishes a lower bound to the value of $\omega_{eff}$ and maintains it in the non-phantom regime, as is also evident from equation \eqref{19}.

Now, from eqns. \eqref{18} and \eqref{19} The Hubble parameter $H$ in terms of redshift $z$ is obtained as
\begin{equation}\label{20}
    H(z)=H_0\biggl[\frac{m(1-n)(1+z)^k -n}{m(1-n)-n}\biggr]^\frac{1}{2(n-1)},
\end{equation}
where $H_0$ is the Hubble value at $z=0$.

 \section{Data Interpretation}
\label{section 3}
This section covers the methods and the range of observational samples used to limit the parameters $H_0$, $m$, $n$, and $k$ of the cosmological model under consideration. Specifically, the posterior distribution of the parameters is obtained via statistical analysis using the Markov Chain Monte Carlo (MCMC) method. The $emcee$ module in Python is used to perform the data analysis portion.\\
The probability function $\mathcal{L}\propto exp(-\chi^{2}/2)$, is used to maximize the best fit of the parameters, with $\chi^{2}$ representing the pseudo-Chi-squared function \cite{hobson2010bayesian}. Further information regarding the $\chi^{2}$ function for different date samples is covered in the following subsections. The MCMC plot displays the $1-D$ curves for each model parameter derived by marginalizing over the other parameters, with a thick-line curve indicating the best-fit value. The panels are on the diagonal in the corner. The off-diagonal panels display $2-D$ projections of the posterior probability distributions for every pair of parameters, along with contours to identify the regions classified as $ 1-\sigma$ (black) and $ 2-\sigma$ (gray).\\

\subsection{Hubble Data}
\begin{figure}[H]
\includegraphics[scale=0.47]{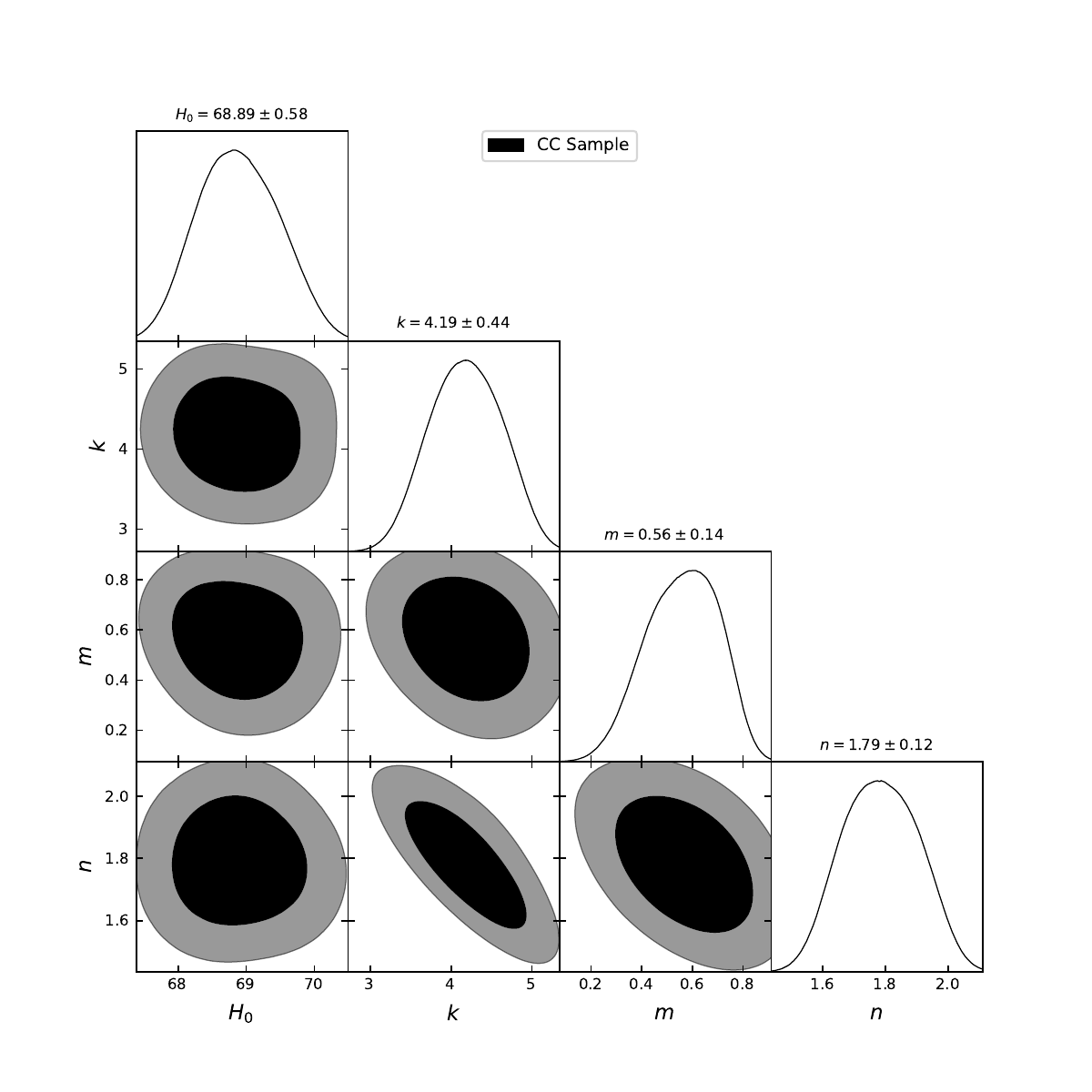}
\caption{\justifying \small{Plot of the MCMC confidence contour obtained after constraining the $f(T)$ power law model with the Hubble dataset. The black-shaded regions represent the $1-\sigma$ CL, and the gray-shaded regions represent the $2-\sigma$ CL. The constraint values for the parameters are presented at the
 $1- \sigma$ CL.}}
\label{III}
\end{figure}

Estimating the expansion rate accurately as a function of cosmic time is difficult. The cosmic chronometers ($CC$) approach is a different and potentially useful technique that takes advantage of the fact that the expansion rate may be written as $H(z) = \Dot{a}/a = -[1/(1 + z)]dz/dt$. The differential age progression of the universe $\Delta t$ at a given redshift interval $\Delta z$ is the only quantity to be measured, since the quantity $\Delta z$ is derived from high-accuracy spectroscopic surveys. Using $\Delta z/\Delta t$, we may approximately calculate the value of $dz/dt$.
To estimate the model parameters, we take 31 data points from the $H(z)$ datasets, which are extracted from the differential age (DA) approach in the redshift range $0.07<z<2.42$. The entire list of this sample is provided collectively in \cite{singirikonda2020model}. To deduce the model parameters, the Chi-square function is given as

\begin{equation}\label{21}
    \chi^2_{CC}=\sum_{i=1}^
    {31} \bigg[\frac{H_{i}^{th}(\Theta_{s},z_{i})-H_{i}^{obs}(z_i)}{\sigma_{H(z_i)}}\biggr]^2,
\end{equation}
where $H^{th}$ and $H^{obs}$ are the theoretical and observed values of the Hubble parameter, respectively. $\Theta_s=(H_0,m,n,k)$ is the
 cosmological background parameter space. $\sigma_{H(z_i)}$ is the standard deviation of the $i^{th}$ point. In Fig. \eqref{I a}, we display the Hubble parameter profile for the CC sample along with the $\Lambda$CDM behavior and Fig. \eqref{I b} shows how the model and the standard $\Lambda$CDM paradigm differ from each other.
 In our MCMC analysis, we used 100 walkers and 1000 steps to determine the fitting results. $1-\sigma$ and $2-\sigma$ confidence level (CL) contour plots are presented in Fig. \eqref{III}. At low redshift, the model is nearly the same; at high redshift, certain differences with the $\Lambda$CDM paradigm are visible. The marginal values of all the model parameters with the Hubble dataset are given in Table \eqref{tab 2}.
 \begin{widetext}
     
\begin{figure}[]
\centering
\begin{subfigure}[b]{0.45\textwidth}
\centering
\includegraphics[width=\textwidth]{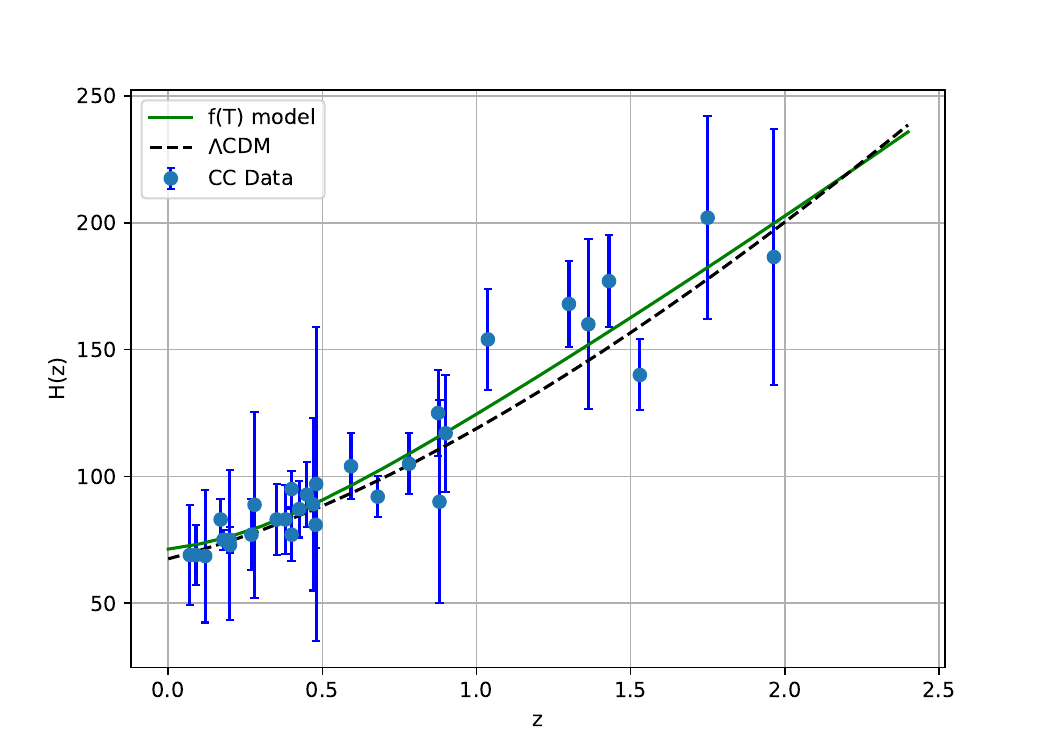}     \caption{\justifying Plot of $H(z)$ versus redshift $(z)$. The given $f(T)$ model (green line), the Hubble parameter profile for $\Lambda$CDM model (black dashed) and the Hubble datasets of 31 data points (blue dots) with
 their corresponding error bars. }      \label{I a}
\end{subfigure}
\hfill
\begin{subfigure}[b]{0.45\textwidth}
        \centering
        \includegraphics[width=\textwidth]{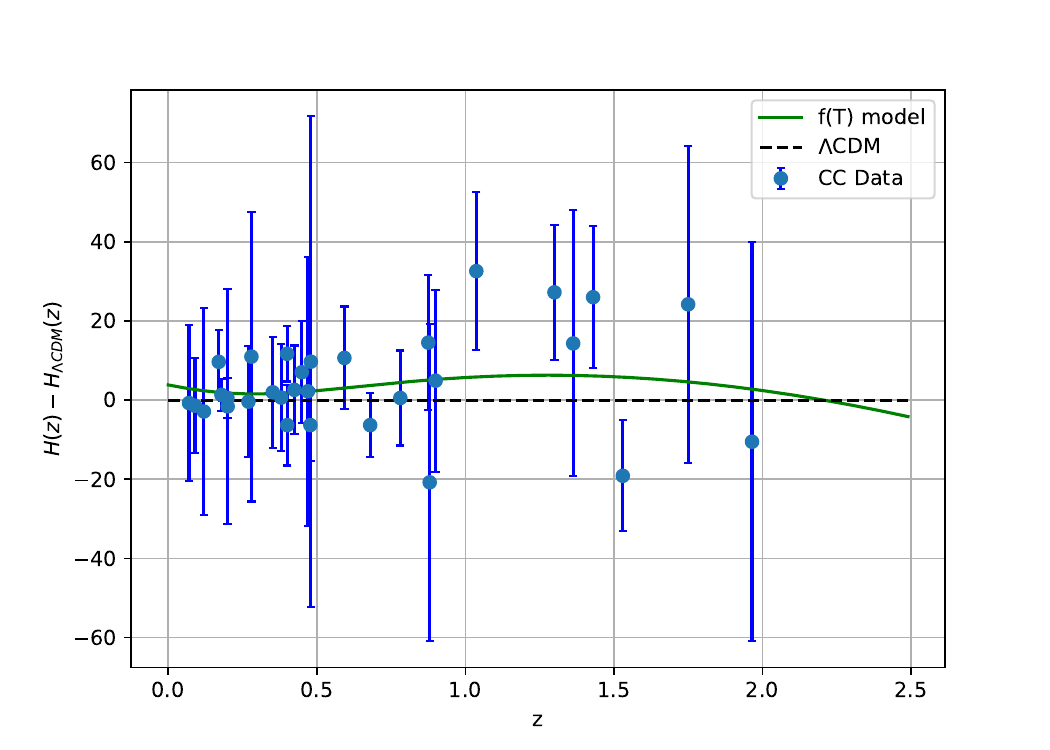}
\caption{\justifying Plot of the relative difference between the $f(T)$ model (green line) and the $\Lambda$CDM model (black dashed) against the 31 Hubble measurements (blue dots),
 along with their corresponding error bars.} \label{I b}
    \end{subfigure}
    \caption{}
    \end{figure}

    \begin{figure}[]
\centering
\begin{subfigure}[b]{0.45\textwidth}
\centering
\includegraphics[width=\textwidth]{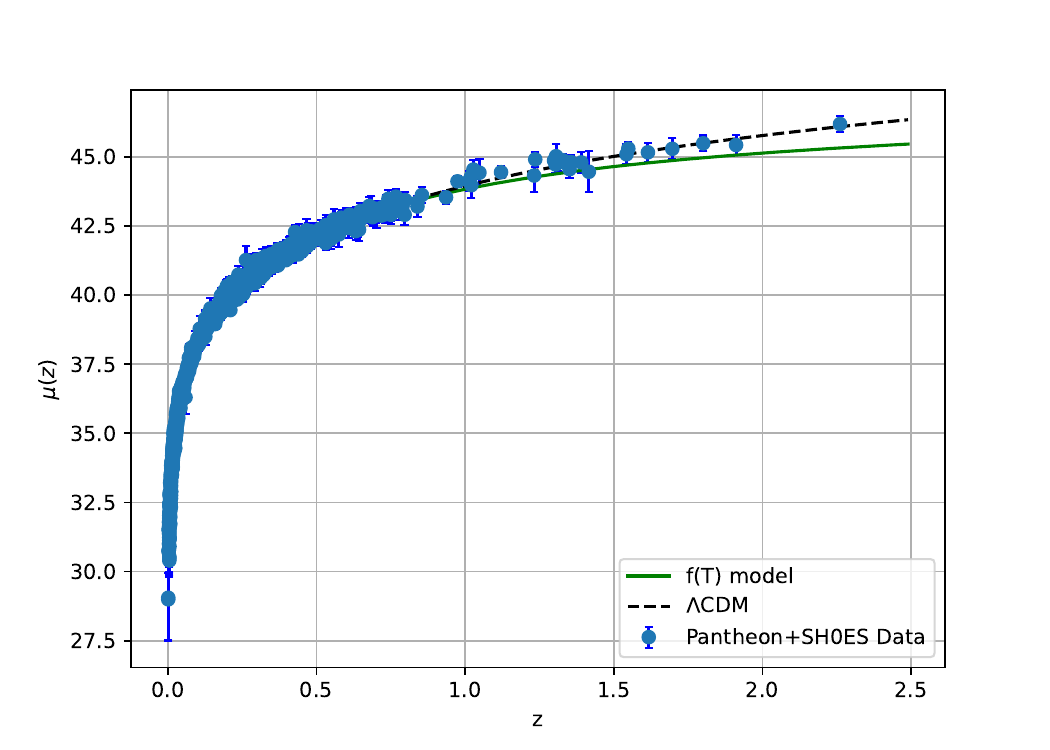}     \caption{\justifying Plot of distance modulus of the $f(T)$ model (green line) and the $\Lambda$CDM model (black dashed) against the 1701 Pantheon+SH0ES datasets (blue dots)
 along with their corresponding error bars.}      \label{II a}
\end{subfigure}
\hfill
\begin{subfigure}[b]{0.45\textwidth}
        \centering
        \includegraphics[width=\textwidth]{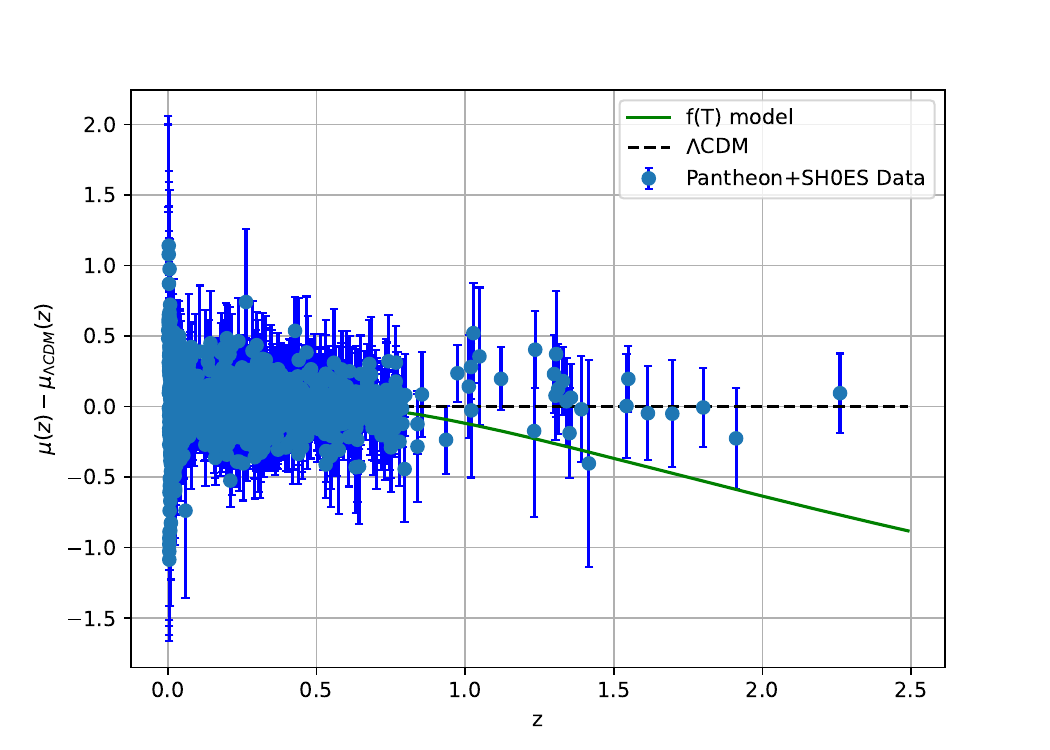}
\caption{\justifying Plot of the relative difference of distance modulus of the $f(T)$ model (green line) and the $\Lambda$CDM model (black dashed) against the 1701 Pantheon+SH0ES datasets (blue dots)
 along with their corresponding error bars.} \label{II b}
    \end{subfigure}
    \caption{}
    \end{figure}
\end{widetext}

\subsection{Pantheon+SH0ES Data}
The discovery of cosmic accelerated expansion has been greatly aided by observations of Type Ia supernovae (SNeIa). SNeIa has proven to be among the most effective methods for examining the characteristics of the elements driving the rapid evolution of the universe. Numerous SNeIa compilations, including joint light-curve analysis (JLA), $Pantheon$, $Pantheon +$, Union, Union 2, and Union 2.1 \cite{kowalski2008improved,amanullah2010spectra,suzuki2012hubble,betoule2014improved,scolnic2018complete}, have been released in recent years. A total of 1701 light curves of 1550 different Type Ia supernovae (SNeIa) with redshifts ranging from $z=0.00122$ to $2.2613$ make up the $Pantheon+SH0ES$ sample collection. By comparing the observed and theoretical values of the distance modulus, the model parameters must be fitted. The theoretical distance modulus $ \mu^{th}_{i}$ can be expressed as
\begin{equation}\label{22}
    \mu^{th}_{i}(z,\theta)=5\log D_l(z,\theta)+25,
\end{equation}
where $D_l$ is the dimensionless luminosity distance defined as,
\begin{equation}\label{23}
D_l(z,\theta)=(1+z)\int^z_0 \frac{d\Bar{z}}{H(\Bar{z})}.
\end{equation}

Now, the Chi-square function is defined as:
\begin{equation}\label{24}
    \chi^2_{SN}(z,\theta)=\sum_{i,j=1}^{1701} \nabla\mu_{i}(C^{-1}_{SN})_{ij} \nabla\mu_{j},
\end{equation}

where $\nabla\mu_i=\mu^{th}_i(z,\theta)-\mu^{obs}_i$ is the difference between the
theoretical and observational distance moduli. $\mu^{obs}_i$ is the observed distance modulus, $\theta$ is the parameter space, and $C_{SN}$ is the covariance matrix \cite{scolnic2022pantheon+}.
%\begin{equation}\label{24}
%\chi^2_{SN}(z,\theta)=\sum_{i=1}^{1701}\bigg[\frac{\mu_{th}(z_i,\theta)-\mu_{obs}(z_i)}{\sigma^2_{\mu}(z_i)}\bigg]^2,
%\end{equation}where $\sigma^2_{\mu}(z_i)$ is the standard error in the observed value. 

\begin{figure}[H]
\includegraphics[scale=0.46]{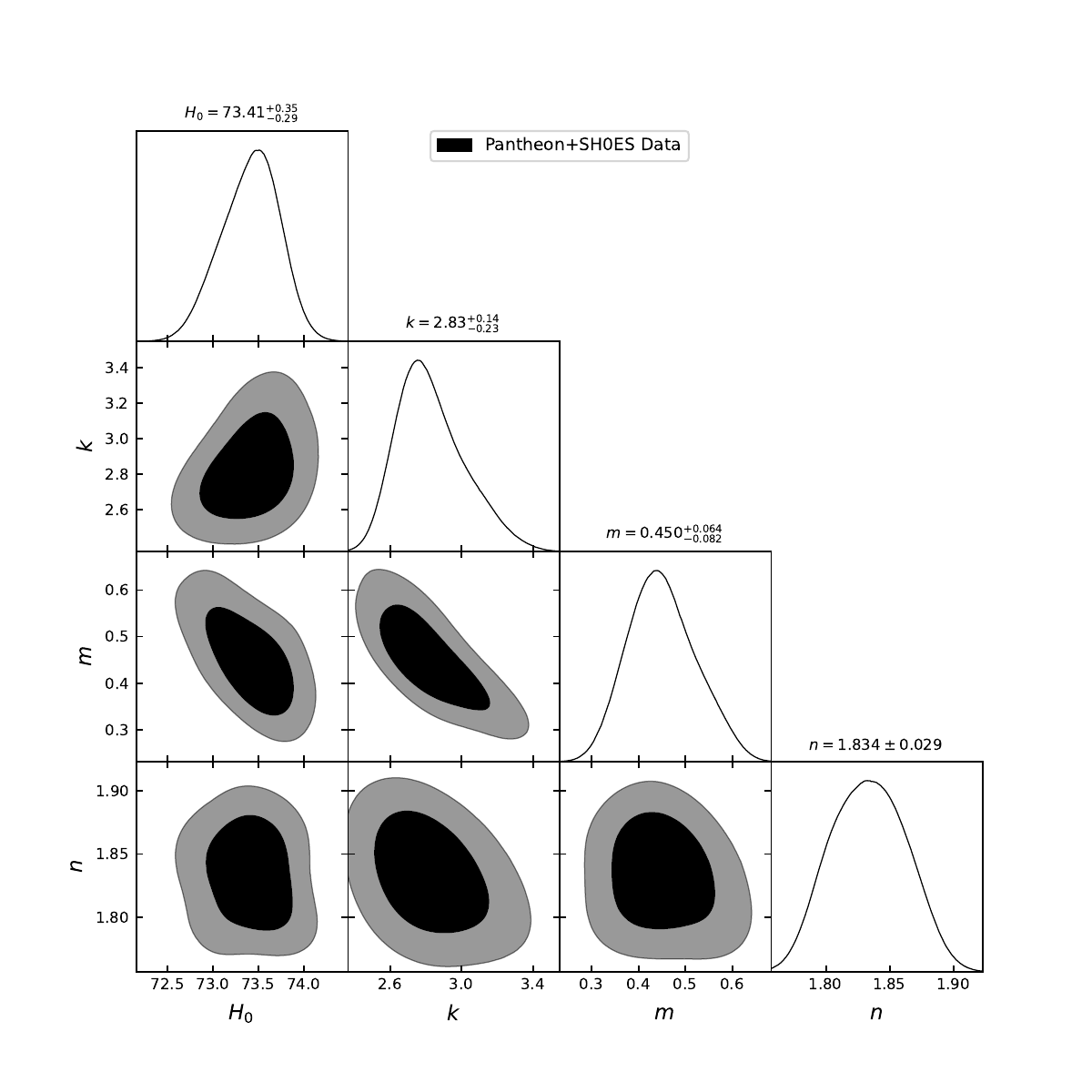}
\caption{\justifying \small{Plot of the MCMC confidence contour obtained after constraining the $f(T)$ power law model with the Pantheon+SH0ES dataset. The black-shaded regions represent the $1-\sigma$ CL, and the gray-shaded regions represent the $2-\sigma$ CL. The constraint values for the parameters are presented at the
 $1- \sigma$ CL.}}
\label{IV}
\end{figure}

We used the same steps and number of walkers in the CC example to run the MCMC code. In Fig. \eqref{II a}, we display the distance modulus parameter, and Fig. \eqref{II b} shows the relative difference profile using the $\Lambda$CDM model and the $Pantheon+SH0ES$ sample. Fig. \eqref{IV} shows the $1-\sigma $ and $2-\sigma$ CL contour plots. One can observe that the model closely matches the $Pantheon+SH0ES$ dataset. The marginal values of all model parameters with the $Pantheon+SH0ES$ dataset are given in Table \eqref{tab 2}.

\subsection{CC+Pantheon+SH0ES Data}
\begin{figure}[H]
\includegraphics[scale=0.45]{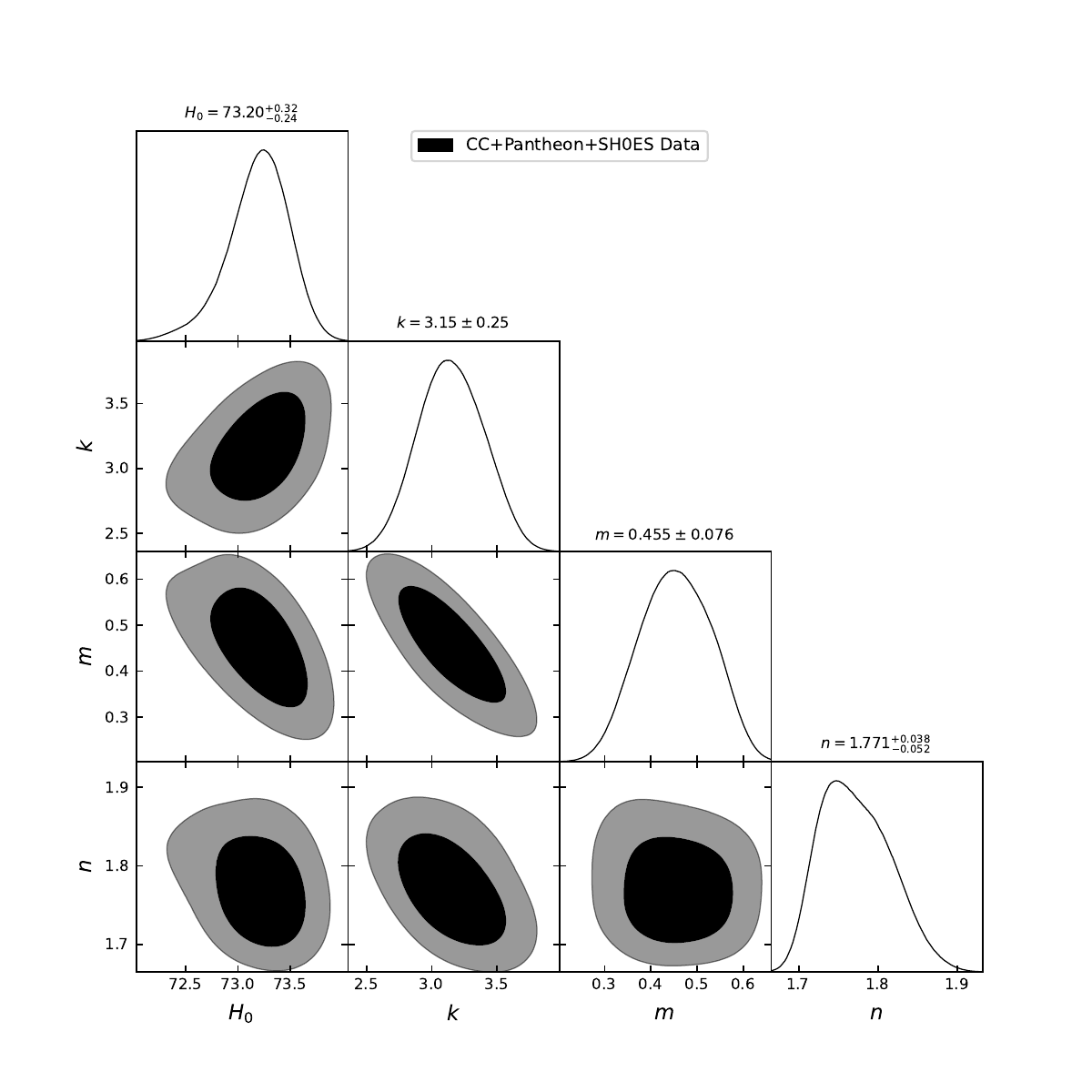}
\caption{\justifying \small{Plot of the MCMC confidence contour obtained after constraining the $f(T)$ power law model with the CC+Pantheon+SH0ES dataset. The black-shaded regions represent the $1-\sigma$ CL, and the gray-shaded regions represent the $2-\sigma$ CL. The constraint values for the parameters are presented at the $1- \sigma$ CL.}}
\label{V}
\end{figure}

By utilizing the CC and Type Ia supernovae samples jointly, the following Chi-square function is employed
\begin{equation}\label{25}
    \chi^2_{CC+SN}=\chi^2_{CC}+\chi^2_{SN}
\end{equation}
Fig. \eqref{V} shows the $1-\sigma $ and $2-\sigma$  CL contour plots. The marginal values of all the model parameters for the combined dataset are given in Table \eqref{tab 2}.\\

%\begin{widetext}
    
\begin{table}[]
\begin{center}
%\adjustbox{width=\textwidth}
{
\begin{tabular}{l c c}
\hline\hline
Datasets& $\chi^2_{min}$ & $\chi^2_{red}$ \\[1ex]
\hline\hline
 CC data & $17.85$ & $0.66$  \\[1ex] 
 $\Lambda$CDM & $17.51$   & $0.65$            \\ \hline
 Pantheon+SH0ES  & $1622.74$ & $0.95$  \\[1ex] 
 $\Lambda$CDM &  $1625.72$  &     $0.96$        \\ \hline
CC+Pantheon+SH0ES  & $1734.8$ & $1.02$ \\[1ex] 
$\Lambda$CDM &  $1735.2$  &     $1.02$        \\
\hline

\end{tabular}
}
\caption{\justifying The corresponding $\chi^2$ values of the models for each sample.}
\label{tab1}
\end{center}
\end{table}
%\end{widetext}

\begin{widetext}

\begin{table}[H]
\centering
%\resizebox{\textwidth}{!}
\begin{tabular}{|c|c|c|c|}
\hline
Parameters  & CC   & Pantheon+SH0ES  & CC+Pantheon+SH0ES \\
\hline
$H_0$ & $68.89\pm 0.58$ & $73.41^{+0.35}_{-0.29}$ & $73.20^{+0.32}_{-0.24}$ \\
\hline
$m$ & $0.56\pm0.14$&$0.45^{+0.064}_{-0.082}$&$0.455\pm0.076$ \\
\hline
$n$ & $1.79 \pm 0.12$ & $1.834 \pm 0.029$ & $1.771^{+0.038}_{-0.052}$\\
\hline
$k$& $4.19 \pm 0.44$ &$2.83^{+0.14}_{-0.23}$ &$3.15\pm 0.25$ \\
\hline
\end{tabular}
\caption{The parametric constraint results obtained via CC, Pantheon+SH0ES, and CC+Pantheon+SH0ES data.}
\label{tab 2}
\end{table}

\end{widetext}

\section{Cosmographic parameters}
\label{section 4}
The deceleration parameter $q$ as a function of the Hubble parameter $H$ is defined as
\begin{equation}\label{26}
    q(z)=-1+ \frac{(1+z)}{H} \frac{dH(z)}{dz},
\end{equation}
The deceleration parameter $q$ plays a crucial role in characterizing the universe's evolution, which shows the acceleration of the universe for  $(q < 0)$ and deceleration for $(q > 0)$ in a given cosmological model. The equation of the deceleration parameter $q$ is obtained as 
\begin{equation}\label{27}
     q(z)=-1 + \frac{km(1+z)^k}{2n+2m(n-1)(1+z)^k}. 
 \end{equation}

\begin{figure}[ht]
\centering
\includegraphics[scale=0.6]{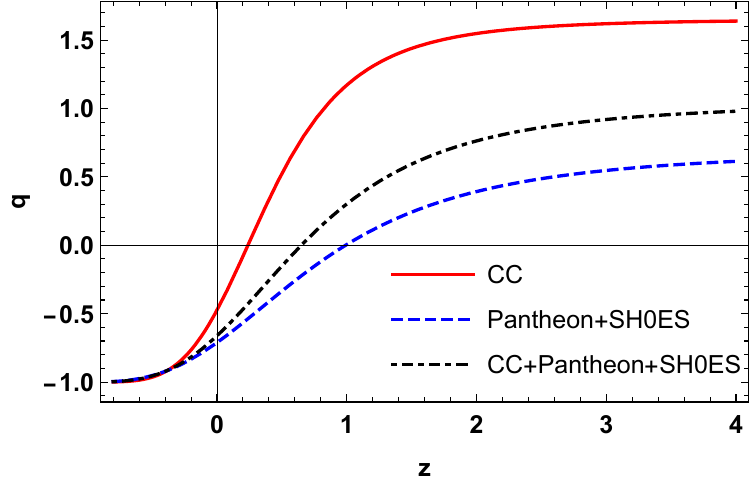}
\caption{\justifying The deceleration parameter $(q)$ versus redshift $(z)$.}
\label{VI}
\end{figure}
 Figure \eqref{VI} illustrates the behavior of the deceleration parameter for the corresponding values of the model parameters $m, n$, and  $k$ specified by $CC$, $Pantheon+SH0ES$, and their compilation. The model demonstrates a transition from a decelerated phase to an accelerated phase. The transition redshifts are determined as $z_t=0.235 \pm 0.026$ $(CC)$ , $z_t = 0.98^{+0.05}_{-0.049}$ $(Pantheon+SHOES)$, and $z_t=0.649 \pm 0.039 $ $(CC+Pantheon+SHOES)$. This finding indicates that the onset of cosmic acceleration occurs at different values of redshift depending on the dataset used. Specifically, $CC$ reveals an earlier transition, whereas $Pantheon+SHOES$ suggests a delayed transition for cosmic acceleration. The present value of the deceleration parameter for each data sample is obtained as $q_0=-0.474^{+0.111}_{-0.122}$, $q_0= -0.711 ^{+0.042}_{-0.061}$, and $q_0= -0.662 ^{+0.066}_{-0.062}$, confirming that the universe is currently in a state of accelerated expansion. These negative values indicate the dominance of dark energy in the present epoch.
 
%It is observed that the transition value of redshift for the $CC$ dataset is found to be $z_t=0.235 \pm 0.026$ and the present value of the deceleration parameter is $q_0=-0.474^{+0.111}_{-0.122}$. In the same way for $CC+Pantheon$ and $CC+Pantheon+SH0ES$ datasets, the transition values of redshift are obtained as $z_t = 0.98^{+0.05}_{-0.049}$ and $z_t=0.649 \pm 0.039$ and  the present value of deceleration parameter $q_0= -0.711 ^{+0.042}_{-0.061}$ and $q_0= -0.662 ^{+0.066}_{-0.062}$ respectively. It is evident from the aforementioned values of $q_0$ for each dataset that, the $CC$ dataset is more compatible with the $\Lambda$CDM model.
 
The total or effective EOS $(\omega_{eff})$ as a function of the Hubble parameter $H$ is given as
\begin{equation}\label{28}
    \omega_{eff}=-1-\frac{2}{3} \frac{\frac{d}{dt}H(z)}{H^2(z)},
\end{equation}
where, $\frac{d}{dt}H(z)=-(1+z)H(z)\frac{d}{dz}H(z)$.
The effective EoS as a whole determines
the relationship between the Hubble parameter and its derivative. The value of $\omega_{eff}$ of the model is obtained from equations \eqref{20} and \eqref{28} as
\begin{equation}\label{29}
    \omega_{eff} = -1 + \frac{km(1+z)^k}{3n +3m(n-1)(1+z)^k}.
\end{equation}

\begin{figure}[ht]
\centering
\includegraphics[scale=0.6]{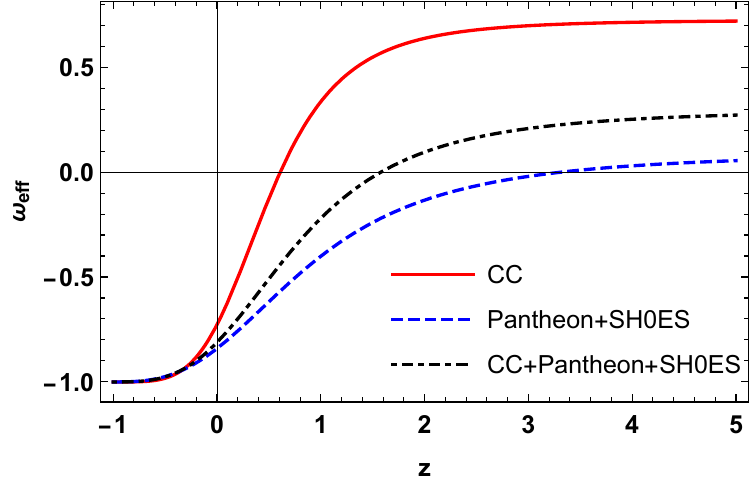}
\caption{\justifying The EoS parameter $(\omega_{eff})$ versus redshift $(z)$ for constrained values of the CC, Pantheon +SH0ES, and CC+Pantheon+SH0ES datasets.}
\label{VII}
\end{figure}

The EOS parameter can indeed be useful in classifying various phases, such as the accelerated and decelerated expansion of the universe. %According to the various phases, a stiff fluid is represented at $\omega = 1$, the radiation-dominated phase is shown by $\omega = 1/3$, and the matter-dominated phase is shown by $\omega = 0$.
Fig. \eqref{VII} displays the evolution trajectory of the effective EoS parameter $\omega_{eff}$. 
%The quintessence era $( -1 < \omega < -1/3)$, the phantom era $ (\omega < -1)$, and the cosmological constant $(\omega = -1) $ are the three conceivable states for the expanding universe.%
The positive range of $\omega_{eff}$ indicates that the model was in the matter-dominant phase, where the rate of expansion slowed. Furthermore, as the universe evolves, DE gradually dominates over matter. This marks the transition to an accelerated expansion phase, with the universe entering a quintessence regime as the effective equation of state (EoS) decreases from -1/3.
%From Fig. \eqref{VII} it is observed that for low redshift the value of the effective EoS parameter fluctuates between $-0.5$ to $-1$ i.e., quintessence dark energy, indicating an acceleration phase which is due to negative effective pressure of the universe explaining the presence of DE.% 
The present value of $\omega_{eff}$, which corresponds to the constrained values of the model parameters, is obtained by fitting the model with the observational data as $\omega_0 = -0.649^{+0.074}_{-0.081}$ ($CC$), $\omega_0 =-0.807^{+0.028}_{-0.041}$ ($Pantheon+SHOES$), and $\omega_0=-0.744 ^{+0.043}_{-0.042}$ ($CC+Pantheon+SHOES$). At present, only  $Pantheon+SH0ES$, but each data sample subsequently strongly agrees with the  $\Lambda$CDM model. 
%value of $\omega_{eff}$ for the $Pantheon+SH0ES$ sample exhibits strong g agreement with the $\Lambda$CDM model.
For the given observational datasets, it is clear that the $\omega_{eff}$ clearly converges to $-1$ but does not cross the phantom divide line ($\omega_{eff}=-1$) in the future, which indicates that the model rules out those extreme scenarios (such as the big rip).
%\begin{equation}r(z)=\frac{2n^2+mn(k^2-3k+4n-4)(1+z)^k+m^2(2+k-2n)(1+k-n)(1+z)^2k}{2\bigl(n+m(n-1)(1+z)^k\bigr)^2}
%\end{equation}
%\begin{equation}s(z)=\frac{m\bigl((k-3)kn+km(3+k-3n)(1+z)^k\bigr)(1+z)^k}{3\bigl(m(3+k-3n)(1+z)^k-3n\bigr)\bigl(n+m(n-1)(1+z)^k\bigr)}
%\end{equation}
%\subsection{Jerk, snap and, lerk parameter}

We attempt to examine additional parameters in $f(T)$ gravity. Expanding the scale factor in the Taylor series concerning the cosmic time \cite{weinberg1972gravitation} indicates a relationship between distance and redshift. The higher-order derivatives of the deceleration parameter, which are referred to as jerk ($j$), snap ($s$), and lerk ($l$) parameters, appear in the Taylor series expansion. These parameters enable us to view the past and future status of the universe.
\begin{equation}\label{30}
    j=(1+z) \frac{dq}{dz} + q(1+2q),
\end{equation}
\begin{equation}\label{31}
    s=-(1+z) \frac{dj}{dz}-q(2+3q),
\end{equation}
\begin{equation}\label{32}
    l=(1+z) \frac{ds}{dz}-s(3+4q).
\end{equation}

  \begin{widetext}
  
\begin{figure}[ht]
\centering
\begin{subfigure}[b]{0.3\textwidth}
\centering
\includegraphics[width=\textwidth]{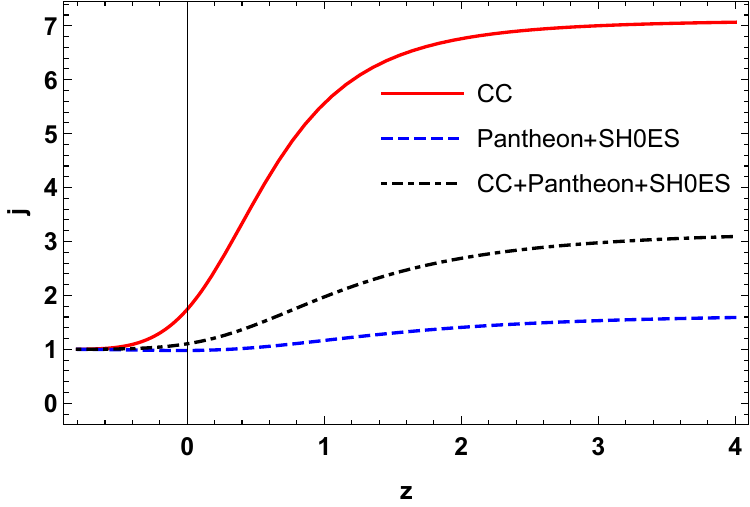}     \caption{}      \label{VIII a}
\end{subfigure}
\hfill
\begin{subfigure}[b]{0.3\textwidth}
        \centering
        \includegraphics[width=\textwidth]{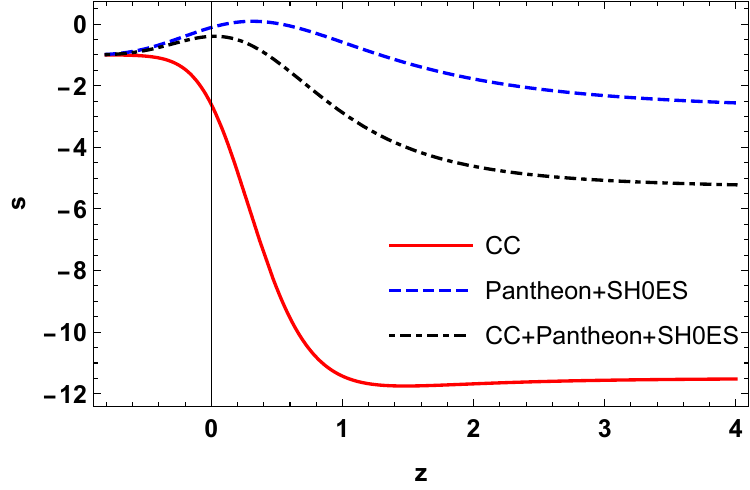}
\caption{} \label{VIII b}
    \end{subfigure}
    \hfill
\begin{subfigure}[b]{0.3\textwidth}    \centering
\includegraphics[width=\textwidth]{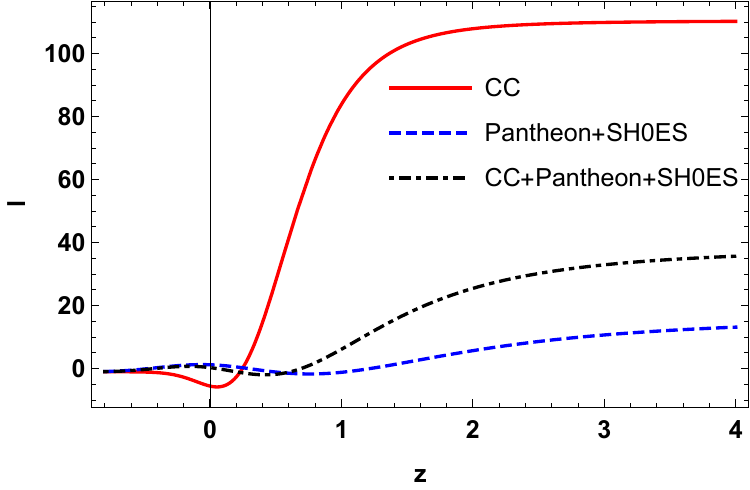}
\caption{} \label{VIII c}
\end{subfigure}
\caption{\justifying The jerk $(j)$, snap $(s)$, and lerk ($l$) parameter versus redshift $(z)$ for constrained values of CC, Pantheon +SH0ES, and CC+Pantheon+SH0ES datasets.}
    \label{fig:main_figure}
\end{figure}
\end{widetext}

From Figs. \eqref{VIII a} and \eqref{VIII c}  the jerk and lerk parameters are observed to have positive decreasing behavior (w.r.t. cosmic time $t$). The dynamics of the universe are controlled by the sign of the jerk parameter, a positive value denotes the existence of a transition phase during which the universe adapts its expansion. Interestingly, for given datasets, the jerk parameter fails to reach unity at $z=0$, i.e., the model deviates from the $\Lambda$CDM model at $z=0$
. The present values of $(j)$ are obtained as $j_{0}=1.741^{+0.368}_{-0.331}, j_{0}= 0.978^{+0.042}_{-0.049}$, and $j_{0}=1.102^{+0.121}_{-0.087}$ for the $CC, Pantheon+SH0ES$, and $CC+Pantheon+SH0ES$ datasets, respectively, but it is observed that the $CC+Pantheon+SH0ES$ datasets have the minimum error of $0.01$ compared with the others; thus at the present interval, the model is more compatible with the standard $\Lambda$CDM model for the $CC+Pantheon+SH0ES$ dataset, and will be exhibited for every dataset in the future. To distinguish between the behavior of a cosmological constant and an evolving dark energy term, the snap parameter plays a crucial role. Fig. \eqref{VIII b} explains the accelerated expansion, which indicates negative behavior and approaches $-1$ late. This demonstrates how late-time acceleration can be observed in a strictly geometrical way under certain modified gravity conditions, such as teleparallel gravity. At the current epoch, the values of the snap parameter are obtained as $s_{0}=-2.606^{+1.074}_{-1.170}$, $s_{0}=-0.110^{+1.071}_{-1.170}$, and $s_{0}=-0.399^{+.130}_{-0.260}$ for the $CC, Pantheon+SH0ES$, and $CC+Pantheon+SH0ES$ datasets, respectively.

\section{\textit{Om} diagnostics}
\label{section 5}
$Om$ is a tool for diagnosis that allows one to distinguish between $\Lambda$CDM with or without density and dynamic DE models. With positive slopes, i.e., ($\omega<-1$), indicating a phantom-type model, and negative slopes, i.e., ($\omega>-1$), showing a quintessence-type model, its consistent behavior, i.e., ($\omega=-1$), suggests that DE is a cosmological constant ($\Lambda$CDM) \cite{sahni2008two}. The $Om(z)$ diagnostic for a spatially flat universe is defined as
\begin{equation}\label{33}
    Om(z)= \frac{E^2(z) -1}{(1+z)^3 -1},
\end{equation}
where $E(z)=H(z)/H_0$.\\ 
\begin{figure}[]
\centering
\includegraphics[scale=0.6]{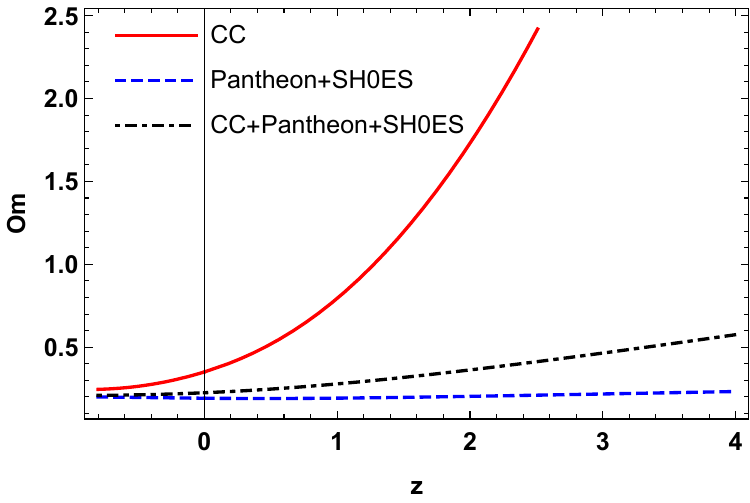}
\caption{\justifying $Om$(z) diagnostic versus redshift ($z$) for constrained values of CC, Pantheon +SH0ES, and CC+Pantheon+SH0ES datasets.}
\label{fig IX}
\end{figure}

Fig. \eqref{fig IX} depicts the behavior of the dark energy model for different datasets. The value of $Om$(z) increases with increasing $z$, which is observed for the $CC$ and $CC+Pantheon+SH0ES$ datasets. The model shows phantom-type behavior as the graph shows an increasing trend, i.e., a positive slope ($\omega<-1$) for the $CC$ and $CC+Pantheon+SH0ES$ datasets. Conversely, for the $Pantheon+SHOES$ dataset, the model remains consistent with the $\Lambda$CDM model.

\begin{widetext}

\begin{table}[ht]
\centering
\begin{center}
%\adjustbox{width=\textwidth}
{\begin{tabular}{l c c c c c c c}\hline\hline
Datasets& $q_0$ & $z_{t}$ & $\omega_0$ & $j_{0}$ & $s_0$ & $l_0$ &  \\
\hline\hline
 CC data & $-0.474^{+0.111}_{-0.122}$ & $0.235 \pm 0.026$   & $ -0.649 ^{+0.074} _{-0.081}$ & $1.741^{+0.368}_{-0.331}$ & $-2.606 ^{+1.074}_{-1.170}$ & $-5.536^{+0.334}_{-2.217}$  &  \\[1ex] 
 Pantheon+SH0ES  & $ -0.711 ^{+0.042}_ {-0.061}$ & $0.98 ^{+0.05}_{-0.049}$   & $-0.807^{+0.028}_{-0.041}$ & $0.978 ^{+0.042}_{-0.049}$ & $-0.110^{+1.071}_{-1.170}$ & $1.222^{+0.115}_{-0.148}$ &  \\[1ex]
CC+Pantheon+SH0ES  & $-0.662^{+0.066}_{-0.062}$ & $0.649 \pm 0.039$   & $-0.774 ^{+0.043}_{-0.042}$ & $1.102^{+0.121}_{-0.087}$ & $-0.399^{+0.130}_{-0.260}$ & $0.310^{+0.495}_{-0.736}$ &  \\[1ex] 
\hline
\end{tabular}
}
\caption{The current values of cosmological parameters for various datasets.}
\label{tab 3}
\end{center}
\end{table}

\end{widetext}

\section{Conclusion}
\label{section6}
In this article, we offer a comprehensive cosmological framework using the power law model of teleparallel gravity within the context of the FLRW universe. To obtain an exact solution of the field equations, we execute the parameterization of the effective EoS as per the \cite{mukherjee2016acceleration}. The optimal results are obtained via the $\chi^2$ minimization technique to get the best-fit values of the model parameters $m,n$, and $k$. To do so, we utilize the dataset samples namely,  $CC, Pantheon+SH0ES$ and perform a joint analysis of $CC$ and $Pantheon+SH0ES$. Table \eqref{tab 2} displays the constrained values of the parameters together with the $1-\sigma$ confidence error. Table \eqref{tab 3} shows the best-fit values of cosmological parameters at the present intervals. We obtain the various cosmological parameters, such as the deceleration parameter, effective EoS parameter, $Om$ diagnostic, jerk, snap, and lerk parameters.

Various observational studies have reported different values of the Hubble constant $H_0$. In modern cosmology, $H_0$-tension arises as a new problem, and we can observe a checklist for different values of $H_0$ \cite{verma2022correction,henning2018measurements}. Therefore, we aspire to constrain $H_0$ in the $f(T)$ gravity framework. For the MCMC analysis, the constrained value of $H_0$ for the $CC$ sample is $H_0 = 68.89\pm 0.58$  $kms^{-1} Mpc^{-1}$, and those for the $Pantheon+SH0ES$ and $CC+Pantheon+SH0ES$ samples are $H_0=73.41^{+0.35}_{-0.29}$ $kms^{-1} Mpc^{-1}$ and $H_0=73.20^{+0.32}_{-0.24}$ $kms^{-1} Mpc^{-1}$ respectively.
The behavior of the deceleration parameter in Fig. \eqref{VI} shows that at a high redshift, the model's accelerated expansion rate increases for the $CC+Pantheon$ sample compared with other samples. In contrast, the model transits into the de Sitter model for every dataset at a low redshift. The present value deceleration parameter for the $CC$ sample is $q_0 \approx -0.474$, which represents a similar cosmic acceleration as the $\Lambda$CDM model. $CC$ shows an early transition, and the $Pantheon+SH0ES$ sample shows a delayed transition to the $\Lambda$CDM model. The model shows a $z_t \approx 0.649$ transition value for $CC+Pantheon+SH0ES$, which is nearly similar to that of the $\Lambda$CDM. Overall, the model shows a transition from initial deceleration to late-time acceleration; this result is in good agreement with \cite{arora2021constraining,myrzakulov2023constrained}. Moving on, Fig. \eqref{VII} explains the negative behavior of the effective EoS. At the current epoch, the models have $\omega_0=- 0.649^{+0.074}_{-0.081}$, $\omega_0 =-0.807^{+0.028}_{-0.041}$, and $\omega_0=-0.744 ^{+0.043}_{-0.042}$ for the given observational dataset samples. Irrespective of each dataset, the model remains in the quintessence era at low redshift. That is, it does not cross the phantom-divide line ($\omega=-1$), which means that in the future, the model will approach $-1$. In another way,  the model so on led to the Einstein-de Sitter model, and our results align well with previously published works \cite{chevallier2001accelerating,arora2021constraining,hazarika2024f}. Moreover, the jerk parameter takes the value of $j_{0}=1.741^{+0.368}_{-0.331}$, which explains the hyperaccelerated expansion for the $CC$ dataset meanwhile, the same model depicts the values $j_{0}= 0.978^{+0.042}_{-0.049}$ and $j_{0}=1.102^{+0.121}_{-0.087}$, which explains the minor and more gradual shift in acceleration for the $Pantheon+SH0ES$ and joint analysis of the $CC$ and $Pantheon+SH0ES$ datasets respectively, as per Fig. \eqref{VIII a}. A noteworthy deviation is observed in the numerical value of the snap parameter $s(z)$ at high and low redshifts for each data sample. Fig. \eqref{VIII b} shows the lower value of $s(z)$ at a high redshift. The present value of this parameter is $s_{0}=-2.606^{+1.074}_{-1.170}$, which indicates a high negative deviation for $CC$, whereas the $Pantheon+SH0ES$ and $CC + Pantheon+SH0ES$ datasets show  $s_{0}=-0.110^{+1.071}_{-1.170}$, and $s_{0}=-0.399^{+.130}_{-0.260}$ respectively, which indicates a low negative deviation from $\Lambda$CDM behavior \cite{bouali2023data,mandal2020accelerating}. Fig. \eqref{fig IX} clearly shows that the occurrence of $Om$(z) at late times is consistent with the breaking down of dark energy theory, and this model is consistent with the standard $\Lambda$CDM model for the $Pantheon+SH0ES$ dataset \cite{sahni2008two,shahalam2015om}.\\
To evaluate whether our model provides a better fit than the $\Lambda$CDM model across various datasets, we computed the minimum chi-square ($\chi^2_{\text{min}}$) and the reduced chi-square ($\chi^2_{\text{red}}$), as shown in Tab. \eqref{tab1}.\\
In conclusion, the power law model of the $f(T)$ gravity is consistent with the observational data and may serve as a potential description of dark energy. Additionally, from both statistical and cosmological perspectives, the derived $f(T)$ model is more compatible with the $\Lambda$CDM model for the $Pantheon+SH0ES$ and $CC+Pantheon+SH0ES$ datasets than the $CC$ dataset.\\

%\begin{widetext}

%\begin{table}[H]
%\begin{center}
%\label{table1}
%\adjustbox{width=\textwidth}
{
%\begin{tabular}{l c c c c c c c}
%\noalign{\doubleline}
%Datasets& $\chi_{\rm min}^2$ & $\chi_{\rm red}^2$ & $ AIC$ & $\Delta AIC$ & $ BIC$ & $\Delta BIC$ \\
%\hline
%CC data &  $17.85$ & $0.66$ & $25.85$ & $5.77$ & $31.59$ & $8.85$\\ 
%\Lambda$CDM &  $16.071$ & $0.595$ & $20.07$ & $0$ & $22.94$ & $0$\\ \hline
%Pantheon+SH0ES & $1623.6174$ & $0.96$ & $1631.6174$ & $17.70$ & $1653.3732$ & $28.58$ \\
%$\Lambda$CDM &  $1609.9172$ & $0.95$ & $1613.9172$ & $0$ & $1624.7951$ & $0$\\ \hline
%CC+Pantheon+SH0ES & $1633.133$ & $1641.133$ & $1662.97$ \\ 
%$\Lambda$CDM &  $16.071$ & $0.595$ & $20.07$ & $0$ & $22.94$ & $0$\\ \hline

%\end{tabular}
}
%\caption{Summary of the best-fit values of model parameters and statistical analysis using SN and OHD+SN+BAO, including the confidence levels.}
%\label{tab 3}
%\end{center}
%\end{table}

%\end{widetext}

\section*{Data Availability Statement}
There are no new data associated with this article.

%\begin{thebibliography}{90}
\bibliographystyle{rsc}
\bibliography{maintext}

\providecommand*{\mcitethebibliography}{\thebibliography}
\csname @ifundefined\endcsname{endmcitethebibliography}
{\let\endmcitethebibliography\endthebibliography}{}
\begin{mcitethebibliography}{54}
\providecommand*{\natexlab}[1]{#1}
\providecommand*{\mciteSetBstSublistMode}[1]{}
\providecommand*{\mciteSetBstMaxWidthForm}[2]{}
\providecommand*{\mciteBstWouldAddEndPuncttrue}
  {\def\EndOfBibitem{\unskip.}}
\providecommand*{\mciteBstWouldAddEndPunctfalse}
  {\let\EndOfBibitem\relax}
\providecommand*{\mciteSetBstMidEndSepPunct}[3]{}
\providecommand*{\mciteSetBstSublistLabelBeginEnd}[3]{}
\providecommand*{\EndOfBibitem}{}
\mciteSetBstSublistMode{f}
\mciteSetBstMaxWidthForm{subitem}
{(\emph{\alph{mcitesubitemcount}})}
\mciteSetBstSublistLabelBeginEnd{\mcitemaxwidthsubitemform\space}
{\relax}{\relax}

\bibitem[Riess \emph{et~al.}((1998))Riess, Filippenko, Challis, Clocchiatti, Diercks, Garnavich, Gilliland, Hogan, Jha, Kirshner,\emph{et~al.}]{riess1998observational}
A.~G. Riess, A.~V. Filippenko, P.~Challis, A.~Clocchiatti, A.~Diercks, P.~M. Garnavich, R.~L. Gilliland, C.~J. Hogan, S.~Jha, R.~P. Kirshner \emph{et~al.}, \emph{The astronomical journal}, (1998), \textbf{116}, 1009\relax
\mciteBstWouldAddEndPuncttrue
\mciteSetBstMidEndSepPunct{\mcitedefaultmidpunct}
{\mcitedefaultendpunct}{\mcitedefaultseppunct}\relax
\EndOfBibitem
\bibitem[Perlmutter \emph{et~al.}((1999))Perlmutter, Aldering, Goldhaber, Knop, Nugent, Castro, Deustua, Fabbro, Goobar, Groom,\emph{et~al.}]{perlmutter1999measurements}
S.~Perlmutter, G.~Aldering, G.~Goldhaber, R.~A. Knop, P.~Nugent, P.~G. Castro, S.~Deustua, S.~Fabbro, A.~Goobar, D.~E. Groom \emph{et~al.}, \emph{The Astrophysical Journal}, (1999), \textbf{517}, 565\relax
\mciteBstWouldAddEndPuncttrue
\mciteSetBstMidEndSepPunct{\mcitedefaultmidpunct}
{\mcitedefaultendpunct}{\mcitedefaultseppunct}\relax
\EndOfBibitem
\bibitem[Spergel \emph{et~al.}((2003))Spergel, Verde, Peiris, Komatsu, Nolta, Bennett, Halpern, Hinshaw, Jarosik, Kogut,\emph{et~al.}]{spergel2003first}
D.~N. Spergel, L.~Verde, H.~V. Peiris, E.~Komatsu, M.~Nolta, C.~L. Bennett, M.~Halpern, G.~Hinshaw, N.~Jarosik, A.~Kogut \emph{et~al.}, \emph{The Astrophysical Journal Supplement Series}, (2003), \textbf{148}, 175\relax
\mciteBstWouldAddEndPuncttrue
\mciteSetBstMidEndSepPunct{\mcitedefaultmidpunct}
{\mcitedefaultendpunct}{\mcitedefaultseppunct}\relax
\EndOfBibitem
\bibitem[Spergel \emph{et~al.}((2007))Spergel, Bean, Dor{\'e}, Nolta, Bennett, Dunkley, Hinshaw, Jarosik, Komatsu, Page,\emph{et~al.}]{spergel2007three}
D.~N. Spergel, R.~Bean, O.~Dor{\'e}, M.~Nolta, C.~Bennett, J.~Dunkley, G.~Hinshaw, N.~e. Jarosik, E.~Komatsu, L.~Page \emph{et~al.}, \emph{The astrophysical journal supplement series}, (2007), \textbf{170}, 377\relax
\mciteBstWouldAddEndPuncttrue
\mciteSetBstMidEndSepPunct{\mcitedefaultmidpunct}
{\mcitedefaultendpunct}{\mcitedefaultseppunct}\relax
\EndOfBibitem
\bibitem[Tegmark \emph{et~al.}((2004))Tegmark, Strauss, Blanton, Abazajian, Dodelson, Sandvik, Wang, Weinberg, Zehavi, Bahcall,\emph{et~al.}]{tegmark2004cosmological}
M.~Tegmark, M.~A. Strauss, M.~R. Blanton, K.~Abazajian, S.~Dodelson, H.~Sandvik, X.~Wang, D.~H. Weinberg, I.~Zehavi, N.~A. Bahcall \emph{et~al.}, \emph{Physical review D}, (2004), \textbf{69}, 103501\relax
\mciteBstWouldAddEndPuncttrue
\mciteSetBstMidEndSepPunct{\mcitedefaultmidpunct}
{\mcitedefaultendpunct}{\mcitedefaultseppunct}\relax
\EndOfBibitem
\bibitem[Eisenstein \emph{et~al.}((2005))Eisenstein, Zehavi, Hogg, Scoccimarro, Blanton, Nichol, Scranton, Seo, Tegmark, Zheng,\emph{et~al.}]{eisenstein2005detection}
D.~J. Eisenstein, I.~Zehavi, D.~W. Hogg, R.~Scoccimarro, M.~R. Blanton, R.~C. Nichol, R.~Scranton, H.-J. Seo, M.~Tegmark, Z.~Zheng \emph{et~al.}, \emph{The Astrophysical Journal}, (2005), \textbf{633}, 560\relax
\mciteBstWouldAddEndPuncttrue
\mciteSetBstMidEndSepPunct{\mcitedefaultmidpunct}
{\mcitedefaultendpunct}{\mcitedefaultseppunct}\relax
\EndOfBibitem
\bibitem[Caldwell((2002))]{caldwell2002phantom}
R.~R. Caldwell, \emph{Physics Letters B}, (2002), \textbf{545}, 23--29\relax
\mciteBstWouldAddEndPuncttrue
\mciteSetBstMidEndSepPunct{\mcitedefaultmidpunct}
{\mcitedefaultendpunct}{\mcitedefaultseppunct}\relax
\EndOfBibitem
\bibitem[Kamenshchik \emph{et~al.}((2001))Kamenshchik, Moschella, and Pasquier]{kamenshchik2001alternative}
A.~Kamenshchik, U.~Moschella and V.~Pasquier, \emph{Physics Letters B}, (2001), \textbf{511}, 265--268\relax
\mciteBstWouldAddEndPuncttrue
\mciteSetBstMidEndSepPunct{\mcitedefaultmidpunct}
{\mcitedefaultendpunct}{\mcitedefaultseppunct}\relax
\EndOfBibitem
\bibitem[Bento \emph{et~al.}((2002))Bento, Bertolami, and Sen]{bento2002generalized}
M.~C. Bento, O.~Bertolami and A.~A. Sen, \emph{Physical Review D}, (2002), \textbf{66}, 043507\relax
\mciteBstWouldAddEndPuncttrue
\mciteSetBstMidEndSepPunct{\mcitedefaultmidpunct}
{\mcitedefaultendpunct}{\mcitedefaultseppunct}\relax
\EndOfBibitem
\bibitem[Li((2004))]{li2004model}
M.~Li, \emph{Physics Letters B}, (2004), \textbf{603}, 1--5\relax
\mciteBstWouldAddEndPuncttrue
\mciteSetBstMidEndSepPunct{\mcitedefaultmidpunct}
{\mcitedefaultendpunct}{\mcitedefaultseppunct}\relax
\EndOfBibitem
\bibitem[Wei and Cai((2008))]{wei2008cosmological}
H.~Wei and R.-G. Cai, \emph{Physics Letters B}, (2008), \textbf{663}, 1--6\relax
\mciteBstWouldAddEndPuncttrue
\mciteSetBstMidEndSepPunct{\mcitedefaultmidpunct}
{\mcitedefaultendpunct}{\mcitedefaultseppunct}\relax
\EndOfBibitem
\bibitem[Gao \emph{et~al.}((2009))Gao, Wu, Chen, and Shen]{gao2009holographic}
C.~Gao, F.~Wu, X.~Chen and Y.-G. Shen, \emph{Physical Review D—Particles, Fields, Gravitation, and Cosmology}, (2009), \textbf{79}, 043511\relax
\mciteBstWouldAddEndPuncttrue
\mciteSetBstMidEndSepPunct{\mcitedefaultmidpunct}
{\mcitedefaultendpunct}{\mcitedefaultseppunct}\relax
\EndOfBibitem
\bibitem[Copeland \emph{et~al.}((2006))Copeland, Sami, and Tsujikawa]{copeland2006dynamics}
E.~J. Copeland, M.~Sami and S.~Tsujikawa, \emph{International Journal of Modern Physics D}, (2006), \textbf{15}, 1753--1935\relax
\mciteBstWouldAddEndPuncttrue
\mciteSetBstMidEndSepPunct{\mcitedefaultmidpunct}
{\mcitedefaultendpunct}{\mcitedefaultseppunct}\relax
\EndOfBibitem
\bibitem[Paliathanasis \emph{et~al.}((2016))Paliathanasis, Barrow, and Leach]{paliathanasis2016cosmological}
A.~Paliathanasis, J.~D. Barrow and P.~Leach, \emph{Physical Review D}, (2016), \textbf{94}, 023525\relax
\mciteBstWouldAddEndPuncttrue
\mciteSetBstMidEndSepPunct{\mcitedefaultmidpunct}
{\mcitedefaultendpunct}{\mcitedefaultseppunct}\relax
\EndOfBibitem
\bibitem[Sharif and Yousaf((2013))]{sharif2013dynamical}
M.~Sharif and Z.~Yousaf, \emph{Physical Review D—Particles, Fields, Gravitation, and Cosmology}, (2013), \textbf{88}, 024020\relax
\mciteBstWouldAddEndPuncttrue
\mciteSetBstMidEndSepPunct{\mcitedefaultmidpunct}
{\mcitedefaultendpunct}{\mcitedefaultseppunct}\relax
\EndOfBibitem
\bibitem[Kadam \emph{et~al.}((2022))Kadam, Mishra, and Said]{kadam2022teleparallel}
S.~Kadam, B.~Mishra and J.~L. Said, \emph{The European Physical Journal C}, (2022), \textbf{82}, 680\relax
\mciteBstWouldAddEndPuncttrue
\mciteSetBstMidEndSepPunct{\mcitedefaultmidpunct}
{\mcitedefaultendpunct}{\mcitedefaultseppunct}\relax
\EndOfBibitem
\bibitem[Bengochea and Ferraro((2009))]{bengochea2009dark}
G.~R. Bengochea and R.~Ferraro, \emph{Physical Review D—Particles, Fields, Gravitation, and Cosmology}, (2009), \textbf{79}, 124019\relax
\mciteBstWouldAddEndPuncttrue
\mciteSetBstMidEndSepPunct{\mcitedefaultmidpunct}
{\mcitedefaultendpunct}{\mcitedefaultseppunct}\relax
\EndOfBibitem
\bibitem[Arora \emph{et~al.}((2021))Arora, Parida, and Sahoo]{arora2021constraining}
S.~Arora, A.~Parida and P.~Sahoo, \emph{The European Physical Journal C}, (2021), \textbf{81}, 1--7\relax
\mciteBstWouldAddEndPuncttrue
\mciteSetBstMidEndSepPunct{\mcitedefaultmidpunct}
{\mcitedefaultendpunct}{\mcitedefaultseppunct}\relax
\EndOfBibitem
\bibitem[Gadbail \emph{et~al.}((2023))Gadbail, Arora, and Sahoo]{gadbail2023dark}
G.~N. Gadbail, S.~Arora and P.~Sahoo, \emph{Annals of Physics}, (2023), \textbf{451}, 169244\relax
\mciteBstWouldAddEndPuncttrue
\mciteSetBstMidEndSepPunct{\mcitedefaultmidpunct}
{\mcitedefaultendpunct}{\mcitedefaultseppunct}\relax
\EndOfBibitem
\bibitem[Myrzakulov \emph{et~al.}(2023)Myrzakulov, Koussour, Alfedeel, and Abebe]{myrzakulov2023constrained}
N.~Myrzakulov, M.~Koussour, A.~H. Alfedeel and A.~Abebe, \emph{The European Physical Journal Plus}, 2023, \textbf{138}, 852\relax
\mciteBstWouldAddEndPuncttrue
\mciteSetBstMidEndSepPunct{\mcitedefaultmidpunct}
{\mcitedefaultendpunct}{\mcitedefaultseppunct}\relax
\EndOfBibitem
\bibitem[Bamba \emph{et~al.}((2016))Bamba, Nashed, El~Hanafy, and Ibraheem]{bamba2016bounce}
K.~Bamba, G.~Nashed, W.~El~Hanafy and S.~K. Ibraheem, \emph{Physical Review D}, (2016), \textbf{94}, 083513\relax
\mciteBstWouldAddEndPuncttrue
\mciteSetBstMidEndSepPunct{\mcitedefaultmidpunct}
{\mcitedefaultendpunct}{\mcitedefaultseppunct}\relax
\EndOfBibitem
\bibitem[Nunes((2018))]{nunes2018structure}
R.~C. Nunes, \emph{Journal of Cosmology and Astroparticle Physics}, (2018), \textbf{2018}, 052\relax
\mciteBstWouldAddEndPuncttrue
\mciteSetBstMidEndSepPunct{\mcitedefaultmidpunct}
{\mcitedefaultendpunct}{\mcitedefaultseppunct}\relax
\EndOfBibitem
\bibitem[Duchaniya \emph{et~al.}((2024))Duchaniya, Gandhi, and Mishra]{duchaniya2024attractor}
L.~Duchaniya, K.~Gandhi and B.~Mishra, \emph{Physics of the Dark Universe}, (2024), \textbf{44}, 101461\relax
\mciteBstWouldAddEndPuncttrue
\mciteSetBstMidEndSepPunct{\mcitedefaultmidpunct}
{\mcitedefaultendpunct}{\mcitedefaultseppunct}\relax
\EndOfBibitem
\bibitem[Blagojevi{\'c} and Nester((2024))]{blagojevic2024lorentz}
M.~Blagojevi{\'c} and J.~M. Nester, \emph{Physical Review D}, (2024), \textbf{109}, 064034\relax
\mciteBstWouldAddEndPuncttrue
\mciteSetBstMidEndSepPunct{\mcitedefaultmidpunct}
{\mcitedefaultendpunct}{\mcitedefaultseppunct}\relax
\EndOfBibitem
\bibitem[Blagojevi{\'c} and Nester(2020)]{blagojevic2020local}
M.~Blagojevi{\'c} and J.~M. Nester, \emph{Physical Review D}, 2020, \textbf{102}, 064025\relax
\mciteBstWouldAddEndPuncttrue
\mciteSetBstMidEndSepPunct{\mcitedefaultmidpunct}
{\mcitedefaultendpunct}{\mcitedefaultseppunct}\relax
\EndOfBibitem
\bibitem[Ferraro and Fiorini(2015)]{ferraro2015remnant}
R.~Ferraro and F.~Fiorini, \emph{Physical Review D}, 2015, \textbf{91}, 064019\relax
\mciteBstWouldAddEndPuncttrue
\mciteSetBstMidEndSepPunct{\mcitedefaultmidpunct}
{\mcitedefaultendpunct}{\mcitedefaultseppunct}\relax
\EndOfBibitem
\bibitem[Kr{\v{s}}{\v{s}}{\'a}k and Saridakis((2016))]{krvsvsak2016covariant}
M.~Kr{\v{s}}{\v{s}}{\'a}k and E.~N. Saridakis, \emph{Classical and Quantum Gravity}, (2016), \textbf{33}, 115009\relax
\mciteBstWouldAddEndPuncttrue
\mciteSetBstMidEndSepPunct{\mcitedefaultmidpunct}
{\mcitedefaultendpunct}{\mcitedefaultseppunct}\relax
\EndOfBibitem
\bibitem[Golovnev((2021))]{golovnev2021issues}
A.~Golovnev, \emph{Classical and Quantum Gravity}, (2021), \textbf{38}, 197001\relax
\mciteBstWouldAddEndPuncttrue
\mciteSetBstMidEndSepPunct{\mcitedefaultmidpunct}
{\mcitedefaultendpunct}{\mcitedefaultseppunct}\relax
\EndOfBibitem
\bibitem[Bahamonde \emph{et~al.}((2023))Bahamonde, Dialektopoulos, Escamilla-Rivera, Farrugia, Gakis, Hendry, Hohmann, Said, Mifsud, and Di~Valentino]{bahamonde2023teleparallel}
S.~Bahamonde, K.~F. Dialektopoulos, C.~Escamilla-Rivera, G.~Farrugia, V.~Gakis, M.~Hendry, M.~Hohmann, J.~L. Said, J.~Mifsud and E.~Di~Valentino, \emph{Reports on Progress in Physics}, (2023), \textbf{86}, 026901\relax
\mciteBstWouldAddEndPuncttrue
\mciteSetBstMidEndSepPunct{\mcitedefaultmidpunct}
{\mcitedefaultendpunct}{\mcitedefaultseppunct}\relax
\EndOfBibitem
\bibitem[Hu \emph{et~al.}((2023))Hu, Zhao, Ren, Wang, Saridakis, and Cai]{hu2023effective}
Y.-M. Hu, Y.~Zhao, X.~Ren, B.~Wang, E.~N. Saridakis and Y.-F. Cai, \emph{Journal of Cosmology and Astroparticle Physics}, (2023), \textbf{2023}, 060\relax
\mciteBstWouldAddEndPuncttrue
\mciteSetBstMidEndSepPunct{\mcitedefaultmidpunct}
{\mcitedefaultendpunct}{\mcitedefaultseppunct}\relax
\EndOfBibitem
\bibitem[Harko \emph{et~al.}((2014))Harko, Lobo, Otalora, and Saridakis]{harko2014nonminimal}
T.~Harko, F.~S. Lobo, G.~Otalora and E.~N. Saridakis, \emph{Physical Review D}, (2014), \textbf{89}, 124036\relax
\mciteBstWouldAddEndPuncttrue
\mciteSetBstMidEndSepPunct{\mcitedefaultmidpunct}
{\mcitedefaultendpunct}{\mcitedefaultseppunct}\relax
\EndOfBibitem
\bibitem[Tajahmad((2017))]{tajahmad2017studying}
B.~Tajahmad, \emph{The European Physical Journal C}, (2017), \textbf{77}, 510\relax
\mciteBstWouldAddEndPuncttrue
\mciteSetBstMidEndSepPunct{\mcitedefaultmidpunct}
{\mcitedefaultendpunct}{\mcitedefaultseppunct}\relax
\EndOfBibitem
\bibitem[Chen \emph{et~al.}((2024))Chen, Wang, Zu, and Fan]{chen2024prospects}
R.~Chen, Y.-Y. Wang, L.~Zu and Y.-Z. Fan, \emph{Physical Review D}, (2024), \textbf{109}, 024041\relax
\mciteBstWouldAddEndPuncttrue
\mciteSetBstMidEndSepPunct{\mcitedefaultmidpunct}
{\mcitedefaultendpunct}{\mcitedefaultseppunct}\relax
\EndOfBibitem
\bibitem[Mukherjee((2016))]{mukherjee2016acceleration}
A.~Mukherjee, \emph{Monthly Notices of the Royal Astronomical Society}, (2016), \textbf{460}, 273--282\relax
\mciteBstWouldAddEndPuncttrue
\mciteSetBstMidEndSepPunct{\mcitedefaultmidpunct}
{\mcitedefaultendpunct}{\mcitedefaultseppunct}\relax
\EndOfBibitem
\bibitem[Chevallier and Polarski((2001))]{chevallier2001accelerating}
M.~Chevallier and D.~Polarski, \emph{International Journal of Modern Physics D}, (2001), \textbf{10}, 213--223\relax
\mciteBstWouldAddEndPuncttrue
\mciteSetBstMidEndSepPunct{\mcitedefaultmidpunct}
{\mcitedefaultendpunct}{\mcitedefaultseppunct}\relax
\EndOfBibitem
\bibitem[Aldrovandi and Pereira((2012))]{aldrovandi2012teleparallel}
R.~Aldrovandi and J.~G. Pereira, \emph{Teleparallel gravity: an introduction}, Springer Science \& Business Media, (2012), vol. 173\relax
\mciteBstWouldAddEndPuncttrue
\mciteSetBstMidEndSepPunct{\mcitedefaultmidpunct}
{\mcitedefaultendpunct}{\mcitedefaultseppunct}\relax
\EndOfBibitem
\bibitem[Rodrigues \emph{et~al.}(2012)Rodrigues, Houndjo, Saez-Gomez, and Rahaman]{rodrigues2012anisotropic}
M.~Rodrigues, M.~Houndjo, D.~Saez-Gomez and F.~Rahaman, \emph{Physical Review D—Particles, Fields, Gravitation, and Cosmology}, 2012, \textbf{86}, 104059\relax
\mciteBstWouldAddEndPuncttrue
\mciteSetBstMidEndSepPunct{\mcitedefaultmidpunct}
{\mcitedefaultendpunct}{\mcitedefaultseppunct}\relax
\EndOfBibitem
\bibitem[Cai \emph{et~al.}((2016))Cai, Capozziello, De~Laurentis, and Saridakis]{cai201679j6901c}
Y.-F. Cai, S.~Capozziello, M.~De~Laurentis and E.~Saridakis, \emph{Rep. Prog. Phys}, (2016),  106901\relax
\mciteBstWouldAddEndPuncttrue
\mciteSetBstMidEndSepPunct{\mcitedefaultmidpunct}
{\mcitedefaultendpunct}{\mcitedefaultseppunct}\relax
\EndOfBibitem
\bibitem[Hobson((2010))]{hobson2010bayesian}
M.~P. Hobson, \emph{Bayesian methods in cosmology}, Cambridge University Press, (2010)\relax
\mciteBstWouldAddEndPuncttrue
\mciteSetBstMidEndSepPunct{\mcitedefaultmidpunct}
{\mcitedefaultendpunct}{\mcitedefaultseppunct}\relax
\EndOfBibitem
\bibitem[Singirikonda and Desai((2020))]{singirikonda2020model}
H.~Singirikonda and S.~Desai, \emph{The European Physical Journal C}, (2020), \textbf{80}, 694\relax
\mciteBstWouldAddEndPuncttrue
\mciteSetBstMidEndSepPunct{\mcitedefaultmidpunct}
{\mcitedefaultendpunct}{\mcitedefaultseppunct}\relax
\EndOfBibitem
\bibitem[Kowalski \emph{et~al.}((2008))Kowalski, Rubin, Aldering, Agostinho, Amadon, Amanullah, Balland, Barbary, Blanc, Challis,\emph{et~al.}]{kowalski2008improved}
M.~Kowalski, D.~Rubin, G.~Aldering, R.~Agostinho, A.~Amadon, R.~Amanullah, C.~Balland, K.~Barbary, G.~Blanc, P.~J. Challis \emph{et~al.}, \emph{The Astrophysical Journal}, (2008), \textbf{686}, 749\relax
\mciteBstWouldAddEndPuncttrue
\mciteSetBstMidEndSepPunct{\mcitedefaultmidpunct}
{\mcitedefaultendpunct}{\mcitedefaultseppunct}\relax
\EndOfBibitem
\bibitem[Amanullah \emph{et~al.}((2010))Amanullah, Lidman, Rubin, Aldering, Astier, Barbary, Burns, Conley, Dawson, Deustua,\emph{et~al.}]{amanullah2010spectra}
R.~Amanullah, C.~Lidman, D.~Rubin, G.~Aldering, P.~Astier, K.~Barbary, M.~Burns, A.~Conley, K.~Dawson, S.~Deustua \emph{et~al.}, \emph{The Astrophysical Journal}, (2010), \textbf{716}, 712\relax
\mciteBstWouldAddEndPuncttrue
\mciteSetBstMidEndSepPunct{\mcitedefaultmidpunct}
{\mcitedefaultendpunct}{\mcitedefaultseppunct}\relax
\EndOfBibitem
\bibitem[Suzuki \emph{et~al.}((2012))Suzuki, Rubin, Lidman, Aldering, Amanullah, Barbary, Barrientos, Botyanszki, Brodwin, Connolly,\emph{et~al.}]{suzuki2012hubble}
N.~Suzuki, D.~Rubin, C.~Lidman, G.~Aldering, R.~Amanullah, K.~Barbary, L.~Barrientos, J.~Botyanszki, M.~Brodwin, N.~Connolly \emph{et~al.}, \emph{The Astrophysical Journal}, (2012), \textbf{746}, 85\relax
\mciteBstWouldAddEndPuncttrue
\mciteSetBstMidEndSepPunct{\mcitedefaultmidpunct}
{\mcitedefaultendpunct}{\mcitedefaultseppunct}\relax
\EndOfBibitem
\bibitem[Betoule \emph{et~al.}((2014))Betoule, Kessler, Guy, Mosher, Hardin, Biswas, Astier, El-Hage, Konig, Kuhlmann,\emph{et~al.}]{betoule2014improved}
M.~Betoule, R.~Kessler, J.~Guy, J.~Mosher, D.~Hardin, R.~Biswas, P.~Astier, P.~El-Hage, M.~Konig, S.~Kuhlmann \emph{et~al.}, \emph{Astronomy \& Astrophysics}, (2014), \textbf{568}, A22\relax
\mciteBstWouldAddEndPuncttrue
\mciteSetBstMidEndSepPunct{\mcitedefaultmidpunct}
{\mcitedefaultendpunct}{\mcitedefaultseppunct}\relax
\EndOfBibitem
\bibitem[Scolnic \emph{et~al.}((2018))Scolnic, Jones, Rest, Pan, Chornock, Foley, Huber, Kessler, Narayan, Riess,\emph{et~al.}]{scolnic2018complete}
D.~M. Scolnic, D.~Jones, A.~Rest, Y.~Pan, R.~Chornock, R.~Foley, M.~Huber, R.~Kessler, G.~Narayan, A.~Riess \emph{et~al.}, \emph{The Astrophysical Journal}, (2018), \textbf{859}, 101\relax
\mciteBstWouldAddEndPuncttrue
\mciteSetBstMidEndSepPunct{\mcitedefaultmidpunct}
{\mcitedefaultendpunct}{\mcitedefaultseppunct}\relax
\EndOfBibitem
\bibitem[Scolnic \emph{et~al.}(2022)Scolnic, Brout, Carr, Riess, Davis, Dwomoh, Jones, Ali, Charvu, Chen,\emph{et~al.}]{scolnic2022pantheon+}
D.~Scolnic, D.~Brout, A.~Carr, A.~G. Riess, T.~M. Davis, A.~Dwomoh, D.~O. Jones, N.~Ali, P.~Charvu, R.~Chen \emph{et~al.}, \emph{The Astrophysical Journal}, 2022, \textbf{938}, 113\relax
\mciteBstWouldAddEndPuncttrue
\mciteSetBstMidEndSepPunct{\mcitedefaultmidpunct}
{\mcitedefaultendpunct}{\mcitedefaultseppunct}\relax
\EndOfBibitem
\bibitem[Weinberg(1972)]{weinberg1972gravitation}
S.~Weinberg, 1972\relax
\mciteBstWouldAddEndPuncttrue
\mciteSetBstMidEndSepPunct{\mcitedefaultmidpunct}
{\mcitedefaultendpunct}{\mcitedefaultseppunct}\relax
\EndOfBibitem
\bibitem[Sahni \emph{et~al.}((2008))Sahni, Shafieloo, and Starobinsky]{sahni2008two}
V.~Sahni, A.~Shafieloo and A.~A. Starobinsky, \emph{Physical Review D—Particles, Fields, Gravitation, and Cosmology}, (2008), \textbf{78}, 103502\relax
\mciteBstWouldAddEndPuncttrue
\mciteSetBstMidEndSepPunct{\mcitedefaultmidpunct}
{\mcitedefaultendpunct}{\mcitedefaultseppunct}\relax
\EndOfBibitem
\bibitem[Verma \emph{et~al.}((2022))Verma, Kashav, Verma, and Chauhan]{verma2022correction}
R.~Verma, M.~Kashav, S.~Verma and B.~Chauhan, \emph{Progress of Theoretical \& Experimental Physics: PTEP}, (2022), \textbf{2022}, \relax
\mciteBstWouldAddEndPuncttrue
\mciteSetBstMidEndSepPunct{\mcitedefaultmidpunct}
{\mcitedefaultendpunct}{\mcitedefaultseppunct}\relax
\EndOfBibitem
\bibitem[Henning \emph{et~al.}((2018))Henning, Sayre, Reichardt, Ade, Anderson, Austermann, Beall, Bender, Benson, Bleem,\emph{et~al.}]{henning2018measurements}
J.~Henning, J.~Sayre, C.~Reichardt, P.~Ade, A.~Anderson, J.~Austermann, J.~Beall, A.~Bender, B.~Benson, L.~Bleem \emph{et~al.}, \emph{The Astrophysical Journal}, (2018), \textbf{852}, 97\relax
\mciteBstWouldAddEndPuncttrue
\mciteSetBstMidEndSepPunct{\mcitedefaultmidpunct}
{\mcitedefaultendpunct}{\mcitedefaultseppunct}\relax
\EndOfBibitem
\bibitem[Hazarika \emph{et~al.}(2024)Hazarika, Arora, Sahoo, and Harko]{hazarika2024f}
A.~Hazarika, S.~Arora, P.~Sahoo and T.~Harko, \emph{arXiv preprint arXiv:2407.00989}, 2024\relax
\mciteBstWouldAddEndPuncttrue
\mciteSetBstMidEndSepPunct{\mcitedefaultmidpunct}
{\mcitedefaultendpunct}{\mcitedefaultseppunct}\relax
\EndOfBibitem
\bibitem[Bouali \emph{et~al.}(2023)Bouali, Chaudhary, Debnath, Sardar, and Mustafa]{bouali2023data}
A.~Bouali, H.~Chaudhary, U.~Debnath, A.~Sardar and G.~Mustafa, \emph{The European Physical Journal Plus}, 2023, \textbf{138}, 816\relax
\mciteBstWouldAddEndPuncttrue
\mciteSetBstMidEndSepPunct{\mcitedefaultmidpunct}
{\mcitedefaultendpunct}{\mcitedefaultseppunct}\relax
\EndOfBibitem
\bibitem[Mandal \emph{et~al.}(2020)Mandal, Bhattacharjee, Pacif, and Sahoo]{mandal2020accelerating}
S.~Mandal, S.~Bhattacharjee, S.~Pacif and P.~Sahoo, \emph{Physics of the Dark Universe}, 2020, \textbf{28}, 100551\relax
\mciteBstWouldAddEndPuncttrue
\mciteSetBstMidEndSepPunct{\mcitedefaultmidpunct}
{\mcitedefaultendpunct}{\mcitedefaultseppunct}\relax
\EndOfBibitem
\bibitem[Shahalam \emph{et~al.}(2015)Shahalam, Sami, and Agarwal]{shahalam2015om}
M.~Shahalam, S.~Sami and A.~Agarwal, \emph{Monthly Notices of the Royal Astronomical Society}, 2015, \textbf{448}, 2948--2959\relax
\mciteBstWouldAddEndPuncttrue
\mciteSetBstMidEndSepPunct{\mcitedefaultmidpunct}
{\mcitedefaultendpunct}{\mcitedefaultseppunct}\relax
\EndOfBibitem
\end{mcitethebibliography}

%\end{thebibliography}

\end{document}